\documentclass[twocolumn, twocolappendix]{aastex701}
\hypersetup{linkcolor=red,citecolor=blue,filecolor=cyan,urlcolor=magenta}
\usepackage{xspace}
\usepackage{aas_macros}
\usepackage{natbib}
\usepackage{graphicx}
\usepackage{siunitx} 
\usepackage{booktabs} 
\usepackage{tabularx} 
\usepackage{textcomp} 
\usepackage[version=4]{mhchem} 
\usepackage[svgnames,dvipsnames]{xcolor} 
\usepackage{txfonts}
\usepackage{amsmath}
\usepackage{subcaption}
\usepackage{ulem} 
\usepackage{soul}

\newcommand{\solar}[1]{\textit{#1}$_\odot$\xspace}
\newcommand{\wavenumbers}{cm$^{-1}$\xspace}
\renewcommand{\micron}{$\mu$m\xspace}
\renewcommand*\farcs{\ensuremath{\overset{\prime\prime}{.}}}

\newcommand{\chthreeplus}{CH$_3^+$\xspace}
\newcommand{\chplus}{CH$^+$\xspace}
\newcommand{\htwo}{H$_2$\xspace}

\newcommand{\CO}{\ce{^{12}CO}\xspace}

\newcommand{\overviewpaper}{Matsuura et al., 2025, MNRAS, in press\xspace}
\newcommand{\baezetal}{Moraga Baez et al., 2025, under review \xspace}
\newcommand{\clarketal}{Clark et al., in prep\xspace}

\shorttitle{Detection of CH$_3^+$ in the O-rich planetary nebula NGC 6302}
\shortauthors{Bhatt et al.}

\begin{document}

\title{Detection of CH$_3^+$ in the O-rich planetary nebula NGC 6302}

\author{Charmi Bhatt}
\affiliation{Department of Physics and Astronomy, University of Western Ontario, London, Ontario, Canada}
\affiliation{Institute for Earth and Space Exploration, University of Western Ontario, London, Ontario, Canada}
\email{cbhatt7@uwo.ca}

\author{Jan Cami}
\affiliation{Department of Physics and Astronomy, University of Western Ontario, London, Ontario, Canada}
\affiliation{Institute for Earth and Space Exploration, University of Western Ontario, London, Ontario, Canada}
\affiliation{SETI Institute, Mountain View, CA, USA}
\email{jcami@uwo.ca}

\author{Els Peeters}
\affiliation{Department of Physics and Astronomy, University of Western Ontario, London, Ontario, Canada}
\affiliation{Institute for Earth and Space Exploration, University of Western Ontario, London, Ontario, Canada}
\affiliation{SETI Institute, Mountain View, CA, USA}
\email{epeeters@uwo.ca}

\author{Nicholas Clark}
\affiliation{Department of Physics and Astronomy, University of Western Ontario, London, Ontario, Canada}
\email{nclark68@uwo.ca}

\author{Paula Moraga Baez}
\affiliation{Center for Imaging Science, Rochester Institute of Technology, Rochester NY 14623, USA}
\email{pm3822@rit.edu}

\author{Kevin Volk}
\affiliation{Space Telescope Science Institute, 3700 San Martin Drive, Baltimore, MD 21218, USA}
\email{volk@stsci.edu}

\author{G.\ C.\ Sloan}
\affiliation{Space Telescope Science Institute, 3700 San Martin Drive, Baltimore, MD 21218, USA}
\affiliation{Department of Physics and Astronomy, University of North Carolina, Chapel Hill, NC 27599-3255, USA}
\email{gcsloan@stsci.edu}

\author{Joel H. Kastner}
\affiliation{Center for Imaging Science, Rochester Institute of Technology, Rochester NY 14623, USA}
\affiliation{School of Physics and Astronomy,  Rochester Institute of Technology, Rochester NY 14623, USA} 
\affiliation{Laboratory for Multiwavelength Astrophysics,  Rochester Institute of Technology, Rochester NY 14623, USA} 
\email{jhkpci@rit.edu}

\author{Harriet L. Dinerstein}
\affiliation{Department of Astronomy, University of Texas at Austin, Austin, TX 78712, USA}
\email{harriet@astro.as.utexas.edu}

\author{Mikako Matsuura}
\affiliation{Cardiff Hub for Astrophysics Research and Technology (CHART), School of Physics and Astronomy, Cardiff University, The Parade, Cardiff CF24 3AA, UK}
\email{matsuuram@cardiff.ac.uk}

\author{Bruce Balick}
\affiliation{Department of Astronomy, University of Washington, Seattle, WA 98195-1580, USA}
\email{balick@uw.edu}

\author{Kathleen E. Kraemer}
\affiliation{Institute for Scientific Research, Boston College, 140 Commonwealth Avenue, Chestnut Hill, MA 02467, USA}
\email{kathleen.kraemer@bc.edu}

\author{Kay Justtanont}
\affiliation{Chalmers University of Technology, Onsala Space Observatory, S-439 92 Onsala, Sweden}
\email{kay.justtanont@chalmers.se}

\author{Olivia Jones}
\affiliation{UK Astronomy Technology Centre, Royal Observatory, Blackford Hill, Edinburgh EH9 3HJ, UK}
\email{olivia.jones@stfc.ac.uk}

\author{Raghvendra Sahai}
\affiliation{Jet Propulsion Laboratory, 4800 Oak Grove Drive, California Institute of Technology,
Pasadena, CA 91109, USA}
\email{sahai@jpl.nasa.gov}

\author{Isabel Aleman}
\affiliation{Laboratório Nacional de Astrofísica, Rua dos Estados Unidos 154, Bairro das Nações, Itajubá, MG 37504-365, Brazil}
\email{bebel.aleman@gmail.com}

\author{Michael J. Barlow}
\affiliation{Department of Physics and Astronomy, University College London, Gower Street, London WC1E 6BT, UK}
\email{mjb@star.ucl.ac.uk} 

\author{Jeronimo Bernard-Salas}
\affiliation{ACRI-ST, Centre d’Études et de Recherche de Grasse (CERGA), 10 Av. Nicolas Copernic, 06130 Grasse, France}
\affiliation{INCLASS Common Laboratory, 10 Av. Nicolas Copernic, 06130 Grasse, France}
\email{jeronimo.bernard-salas@ias.u-psud.fr}

\author{Joris Blommaert}
\affiliation{Astronomy and Astrophysics Research Group, Department of Physics and Astrophysics, Vrije Universiteit Brussel, Pleinlaan 2, B-1050 Brussels, Belgium}
\email{joris.blommaert@vub.be}


\author{Naomi Hirano}
\affiliation{Academia Sinica Institute of Astronomy and Astrophysics, 11F of Astronomy-Mathematics Building, 
AS/NTU, No.1, Sec. 4, Roosevelt Road, Taipei 106319, Taiwan, R.O.C. }
\email{hirano@asiaa.sinica.edu.tw}

\author{Patrick Kavanagh}
\affiliation{Department of Physics, Maynooth University, Maynooth, County Kildare, Ireland}
\email{paddy.astro.dias@gmail.com}

\author{Francisca Kemper}
\affiliation{Institut de Ciències de l’Espai (ICE, CSIC), Can Magrans, s/n, E-08193 Cerdanyola del Vallès, Barcelona, Spain}
\affiliation{ICREA, Pg. Lluís Companys 23, E-08010 Barcelona, Spain}
\affiliation{Institut d’Estudis Espacials de Catalunya (IEEC), E-08860 Castelldefels, Barcelona, Spain}
\email{ciska.kemper@icrea.cat}

\author{Eric Lagadec}
\affiliation{Université Côte d’Azur, Observatoire de la Côte d’Azur, CNRS, Lagrange, 96 Bd de l’Observatoire, 06300 Nice, France}
\email{eric.lagadec@oca.eu}

\author{J. Martin Laming}
\affiliation{Space Science Division, Naval Research Laboratory, Code 7684, Washington, DC 20375, USA}
\email{j.m.laming.civ@us.navy.mil}

\author{Frank Molster}
\affiliation{Leidse instrumentmakers School, Einsteinweg 61, 2333 CC Leiden, Netherlands}
\email{Molster@Lis.nl}

\author{Hektor Monteiro}
\affiliation{Cardiff Hub for Astrophysics Research and Technology (CHART), School of Physics and Astronomy, Cardiff University, Cardiff CF24 3AA, UK}
\affiliation{Instituto de Física e Química, Universidade Federal de Itajubá, Av. BPS 1303, 37500-903 Itajubá, MG, Brazil}
\email{hmonteiro@unifei.edu.br}

\author{Anita M. S. Richards}
\affiliation{Jodrell Bank Centre for Astrophysics, Department of Physics \& Astronomy, The University of Manchester, Oxford Road, Manchester M13 9PL, UK}
\email{a.m.s.richards@manchester.ac.uk}

\author{N.\ C.\ Sterling}
\affiliation{University of West Georgia, 1601 Maple Street, Carrollton, GA 30118, USA}
\email{nsterlin@westga.edu}

\author{Maryam Torki}
\affiliation{Institut de Ciències de l’Espai (ICE, CSIC), Can Magrans, s/n, E-08193 Cerdanyola del Vallès, Barcelona, Spain}
\email{mtorki@ice.csic.es}

\author{Peter A. M. van Hoof}
\affiliation{Royal Observatory of Belgium, Ringlaan 3, B-1180 Brussels, Belgium}
\email{p.vanhoof@oma.be}

\author{Jeremy R. Walsh}
\affiliation{European Southern Observatory, Karl-Schwarzschild-Strasse 2, D-85748 Garching, Germany}
\email{jwalsh@eso.org}

\author{L. B. F. M. Waters}
\affiliation{Department of Astrophysics/IMAPP, Radboud University, PO Box 9010, 6500 GL Nijmegen, The Netherlands}
\affiliation{SRON Netherlands Institute for Space Research, Niels Bohrweg 4, 2333 CA Leiden, The Netherlands}
\email{rens.waters@ru.nl}

\author{Roger Wesson}
\affiliation{Cardiff Hub for Astrophysics Research and Technology (CHART), School of Physics and Astronomy, Cardiff University, Cardiff CF24 3AA, UK}
\affiliation{Department of Physics and Astronomy, University College London, Gower Street, London WC1E 6BT, UK}
\email{wessonr1@cardiff.ac.uk}

\author{Finnbar Wilson}
\affiliation{Cardiff Hub for Astrophysics Research and Technology (CHART), School of Physics and Astronomy, Cardiff University, Cardiff CF24 3AA, UK}
\email{wilsonf5@cardiff.ac.uk}

\author{Nicholas J. Wright}
\affiliation{Astrophysics Research Centre, Keele University, Newcastle-under-Lyme, ST5 5BG, UK}
\email{n.j.wright@keele.ac.uk}

\author{Albert A. Zijlstra}
\affiliation{Jodrell Bank Centre for Astrophysics, Department of Physics \& Astronomy, The University of Manchester, Oxford Road, Manchester M13 9PL, UK}
\affiliation{School of Mathematical and Physical Sciences, Macquarie University, Sydney, NSW 2109, Australia}
\email{a.zijlstra@manchester.ac.uk}

\begin{abstract}

Planetary nebulae are sites where ejected stellar material evolves into complex molecules, but the precise physical conditions and chemical routes that govern these processes are unclear. The presence of abundant carbon-rich molecules in O-rich environments poses particular challenges. Here we report the first detection of methyl cation (\chthreeplus) in any planetary nebula, observed in the O-rich nebula NGC~6302 using JWST MIRI/MRS observations. \chthreeplus is a key driver of organic chemistry in UV-irradiated environments. Spatially resolved observations reveal that \chthreeplus is co-located with \CO, \htwo, \ion{H}{2}, \ce{HCO+}, and Polycyclic aromatic hydrocarbons (PAHs).  LTE modelling of the \chthreeplus emission yields excitation temperatures of 500-800K in the inner bubble and torus, rising to 1000-2000K in the outer bubble of NGC~6302, with column densities ranging from $\sim$10$^{11}$ to 10$^{13}$ cm$^{-2}$. This detection demonstrates that hydrocarbon radical chemistry must be incorporated into planetary nebulae chemical models. Further near-IR observations are crucial to map different chemical networks operating in these environments. 

\end{abstract}


\keywords{\uat{Astrochemistry}{75}, \uat{Circumstellar Gas}{238}, \uat{Planetary nebulae}{1249}, \uat{Small molecules}{2267}}


\section{Introduction} 

\begin{figure*}
    \centering
    \includegraphics[width=\linewidth]{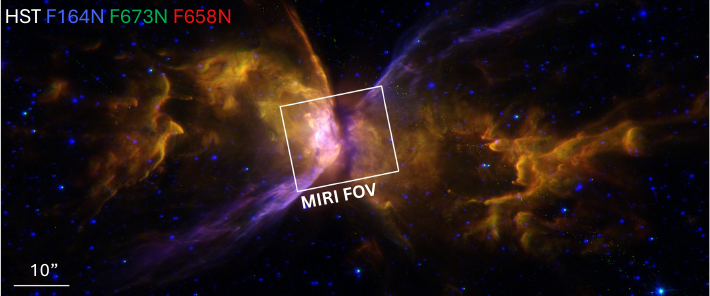}
    \caption{Three-colour composite HST image of NGC 6302, created by Matsuura et al. (2025, MNRAS, submitted) utilizing images from \citet{2022ApJ...927..100K} and \cite{2023ApJ...957...54B}. This image presents a colour overlay of WFC3  observations of NGC 6302, featuring F164N filter in blue, which traces [Fe~{\sc ii}] emission, F656N filter in green, which traces H$\alpha$ emission, and F658N in red, which traces [N~{\sc ii}] emission. The white box at the center represents the MIRI field-of-view for observations described in Sect.~\ref{sec:data}.}
    \label{fig:HST}
\end{figure*}

As of August 2025, more than 330 molecular species have been identified in interstellar and circumstellar environments. A substantial subset of this inventory originates in the circumstellar envelopes of evolved stars, including stars on the asymptotic giant branch (AGB), post-AGB stars, and planetary nebulae \citep[singular: PN; plural: PNe;][]{2022ApJS..259...30M, moleculesinspace}. Elucidating how these molecules form and evolve as local physical conditions change is essential for constraining a wide range of astrochemical and astrophysical processes such as star- and planet formation and the chemical enrichment of galaxies, among others \citep[see][for a review]{2013RvMP...85.1021T}.

During the AGB phase, low- to intermediate-mass ($\sim$\mbox{1--8~\solar{M}}) stars undergo severe mass-loss with mass-loss rates ranging from 10$^{-8}$ to 10$^{-4}$ \solar{M} yr$^{-1}$ \citep{2018A&ARv..26....1H}. These AGB outflows result in the formation of a circumstellar envelope (CSE). As the stellar gas expands radially away from the stellar atmosphere, it cools down, resulting in the formation of molecules and dust. The infrared (IR) spectra of AGB stars indeed show the presence of a rich variety of molecules. More than 90 molecules have been detected in the CSE around carbon-rich stars, and about 27 around oxygen-rich stars  \citep[see][and references there in]{2014A&A...570A..45A, 2018A&A...615A..28D, 2022EPJWC.26500029A, 2022ApJS..259...30M}. Note that the molecular species detected in the CSE of AGB stars are mainly simple molecular species, such as CO, C$_2$H$_2$, HCN, CO$_2$, H$_2$O, etc. Eventually, the star progresses through the post-AGB phase and the effective temperature of the star, \textit{T}$_{\text{eff}}$, increases rapidly with decreasing stellar radii. As \textit{T}$_{\text{eff}}$ $\gtrsim$20,000K, UV radiation from the star ionizes the hydrogen in the previously ejected material and a PN is formed \citep{2022PASP..134b2001K}. UV photons from the star can also lead to dissociation of molecules in the CSE, and not all molecules survive the transition from the AGB to the PN phase. In fact, a sharp drop is seen in the molecular inventory of PN compared to that of the AGB phase \citep{2022EPJWC.26500029A}. Interestingly, spectral features of larger and more complex molecules like Polycyclic Aromatic Hydrocarbons (PAHs) and fullerenes, such as C$_{60}$, C$_{70}$, begin to appear in the post-AGB and PN phase \citep{2005MNRAS.359..383M, 2014MNRAS.439.1472M, 2016JPhCS.728c2011C, Cami:C60-Science, ZhangKwok:proto-PNC60}. 


A key concept in the chemistry of evolved stars is that the chemistry in the AGB outflows depends primarily on the C/O ratio of the stellar photosphere. All stars start their lives as oxygen-rich (C/O $<$ 1) stars, but third dredge-up events in low- and intermediate-mass stars \citep[$\sim$1.5--4 \solar{M}][]{2013A&A...555L...3G} elevate this C/O ratio above unity late in the AGB phase, converting the envelope to carbon-rich. After \htwo, the next molecule to form in the stellar photosphere is carbon monoxide (CO), which is a highly stable molecule. CO locks up all the free carbon or oxygen atoms -- depending on which species is most abundant -- leaving the remaining carbon or oxygen to drive further chemistry. Oxygen-rich stellar envelopes  thus exhibit the presence of oxides (such as \ce{H2O}, \ce{SO}, \ce{AlOH}, etc.),  and crystalline silicates \citep{1999A&A...352..587S, Cami:PhD, 2002A&A...393L...7P, 2010A&A...516A..68K, 2010A&A...516A..69D, 2013A&A...558A.132D, 2018A&A...615A..28D} while carbon-rich stellar envelopes  are characterized by the presence of hydrocarbons, PAHs, fullerenes and amorphous dust \citep{2001ApJ...562..824H, Cami:C60-Science, 2014A&A...570A..45A} 

However, \citet{1991AandA...243L...9Z} presented the first evidence of mixed or dual chemistry in a PN - where the spectrum of a young PN (IRAS 07027-7934) showed the simultaneous presence of strong PAH emission as well as 1.6 GHz OH maser line emission. Using observations by the Infrared Space Observatory, more cases of dual chemistry in PNe were reported by \citet{1998A&A...331L..61W,1998Natur.391..868W} and \citet{2002MNRAS.332..879C}. Furthermore, using Spitzer observations, \citet{2009A&A...495L...5P} showed that out of 26 PNe observed towards the galactic bulge, 21 showed signatures of dual chemistry. A larger survey by \citet{2012ApJ...753..172S} showed that 46\% of the bulge PNe and 24\% of the galactic PNe exhibit dual chemistry. It is not clear whether this dual-chemistry phenomenon arises on the AGB or emerges later as a result of processing during the PN phase. Consequently, our knowledge and understanding of the chemical pathways in late stages of stellar evolution, from the AGB phase to PN, is incomplete. It is important to explore how the inventory of molecules is affected by the dual chemistry, and eventually, what chemical reactions can occur which are not accounted for by strictly C-rich or O-rich chemical pathways. This can potentially provide clues to the formation of other chemical species; for example, the formation of large carbonaceous aromatic molecules like PAHs in an O-rich environment. 


In this paper, we report the unambiguous detection of the methyl cation (\chthreeplus) in the NGC~6302, a small but important carbonaceous species in what is very clearly an O-rich environment \citep[C/0 $\approx$ 0.5,][]{2011MNRAS.418..370W}. While \chthreeplus has long been considered a key species for carbon chemistry in interstellar environments \citep{1977ApJS...34..405B, smith_ion_1992, 1995ApJS...99..565S, 2021FrASS...8..207H}, its first detection outside the Solar System is a much more recent milestone. \chthreeplus was first detected via JWST/MIRI observations of the irradiated proplyd d203-506 in the Orion Bar by \citet{2023Natur.621...56B}. The species is also present at much larger scales in the Orion Bar PDR \citep{2025A&A...696A..99Z} and in the disk of a T-Tauri star \citep{2024PASP..136e4302H}. Our detection in NGC~6302 is the first time this species has been detected in a planetary nebula. 


\section{Observations and data reduction}
\label{sec:data}

We use the JWST MIRI/MRS observations of NGC~6302, which were obtained as part of the JWST GO Cycle 1 program 1742. The pointing was centred on the central region of NGC~6302 at R.A.= 17$^h$:13$^m$;44.3938$^s$, Dec. = -37$^\circ$:06$'$:12.36 (J2000). The observations were taken with the MIRI Medium Resolution Spectrometer (MRS) with Integral Field Unit (IFU) . The data was obtained for the entire MIRI wavelength range, which is 4.9 to 27.9~\micron spread across 4 channels \citep{2015PASP..127..646W, 2023A&A...675A.111A}.  A mosaic of 5 x 5 tiles was used to cover the central part of NGC 6302, which includes the central star, the torus, and the bright ionized line emitting region at the innermost region of the bipolar outflow (see fig.~\ref{fig:HST}). Four-point dithering was used to optimize spatial and spectral sampling throughout the field of view. 

We used the development version of v1.14.0 JWST Calibration Pipeline with versions 11.17.16 and ``jwst\_1202.pmap'' of the Calibration Reference Data System (CRDS) and CRDS context, respectively. All level 1b files were processed through \texttt{Detector1Pipeline} to obtain level 2a (rate) images. The background exposures were subtracted from the science exposures to get background-subtracted level 2a files. These were then processed through \texttt{Spec2Pipeline} (with \texttt{residual\_fringe} step tuned on), followed by \texttt{Spec3Pipeline} to produce spectral cube mosaics of NGC 6302. For details on the observations and data reduction, we refer to \overviewpaper.

\section{The planetary nebula NGC~6302}
\label{sec:target-of-interest}

\begin{figure*}
  \centering

  \gridline{
  \fig{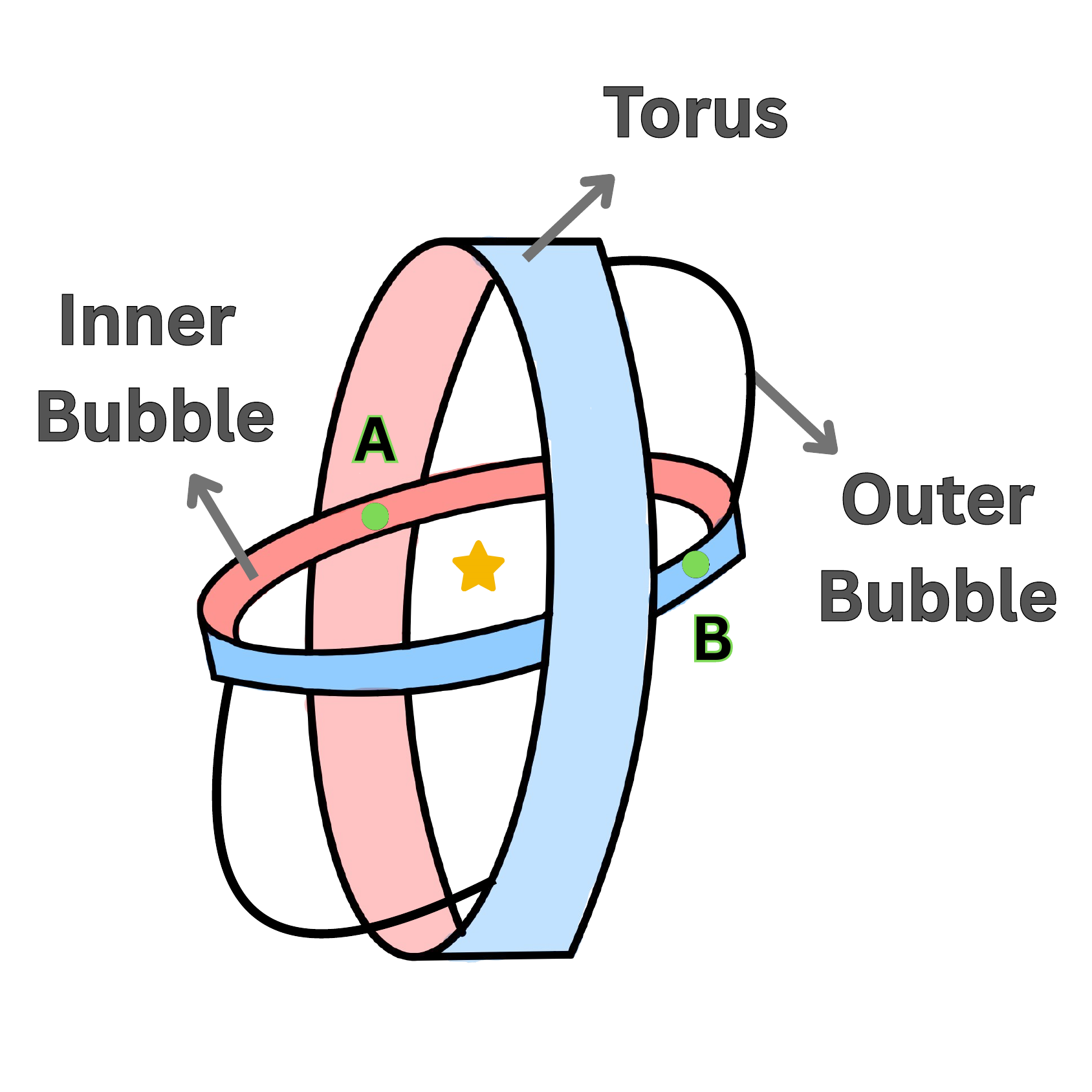}{0.3\textwidth}{}
  \fig{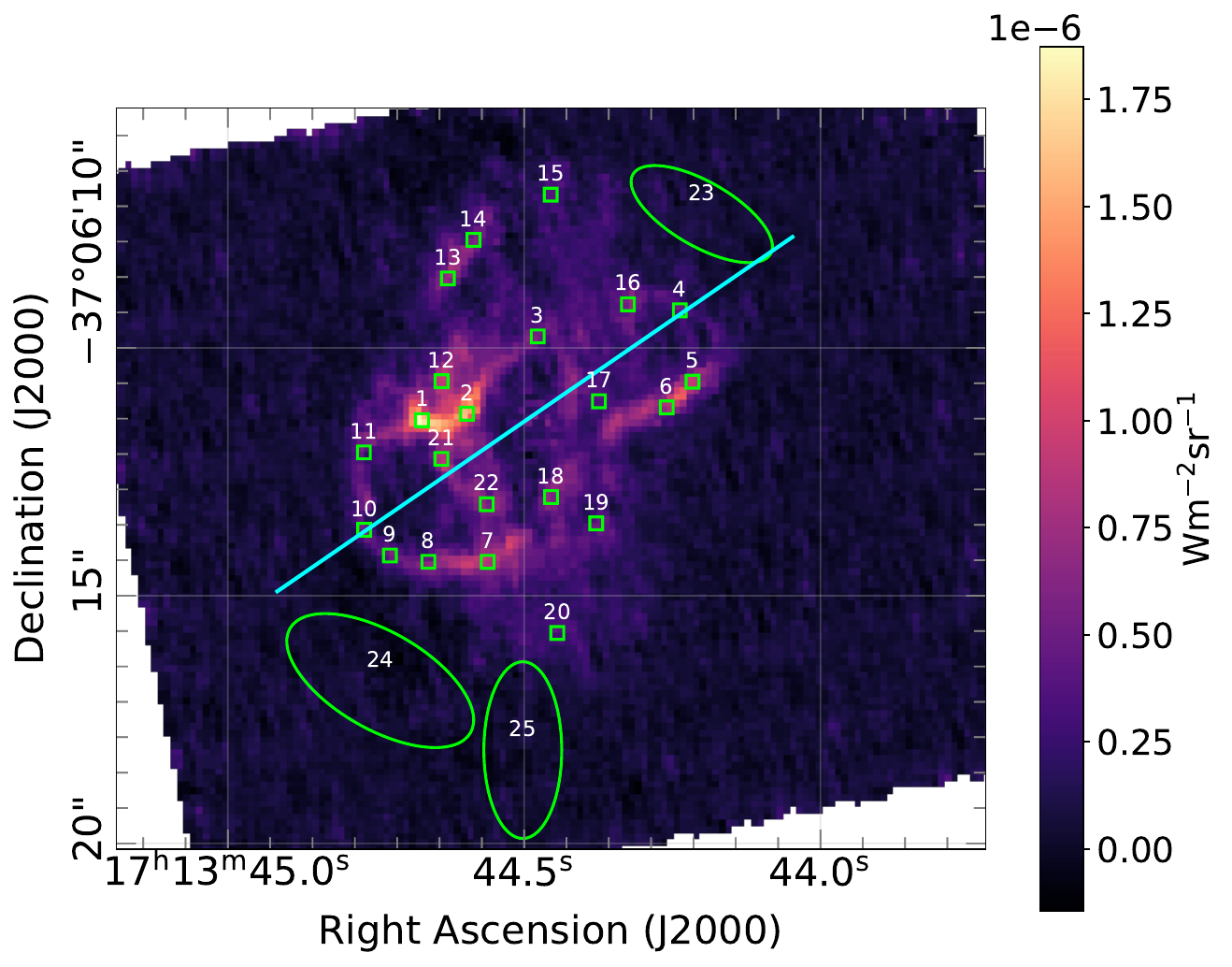}{0.58\textwidth}{}
  }
  \caption{\textit{Left panel:} The cartoon provides a schematic representation of the central region of NGC~6302. Blue and red colours indicate the blueshift and redshift caused by the expansion of the torus and inner bubble from the Earth's perspective. The schematic is based on the results of \citet{2017A&A...597A..27S} and \overviewpaper. The line-of-sight motion of the outer bubble is not constrained and thus is represented by arcs. Points A and B are two bright spots on \chthreeplus integrated surface brightness map. \textit{Right panel:}Integrated surface brightness map of \chthreeplus, obtained by integrating the surface brightness over the 7.13--7.2~\micron wavelength range at each pixel. The green boxes and ellipses represent the 25 apertures where we analyzed the \chthreeplus emission in more detail. Aperture 1-22 are 2$\times$2 pixels, which is 0\farcs26 $\times$ 0\farcs26. Apertures 23-25 are larger elliptical regions with areas of 3\farcs22$^2$, 6\farcs52$^2$, and 4\farcs36$^2$, respectively. The cyan line represents the cut used for making Fig.~\ref{fig:crosscuts}. }
\label{fig:cartoon_and_apertures}
\end{figure*}

\begin{figure}
    \centering
    \includegraphics[width=\linewidth]{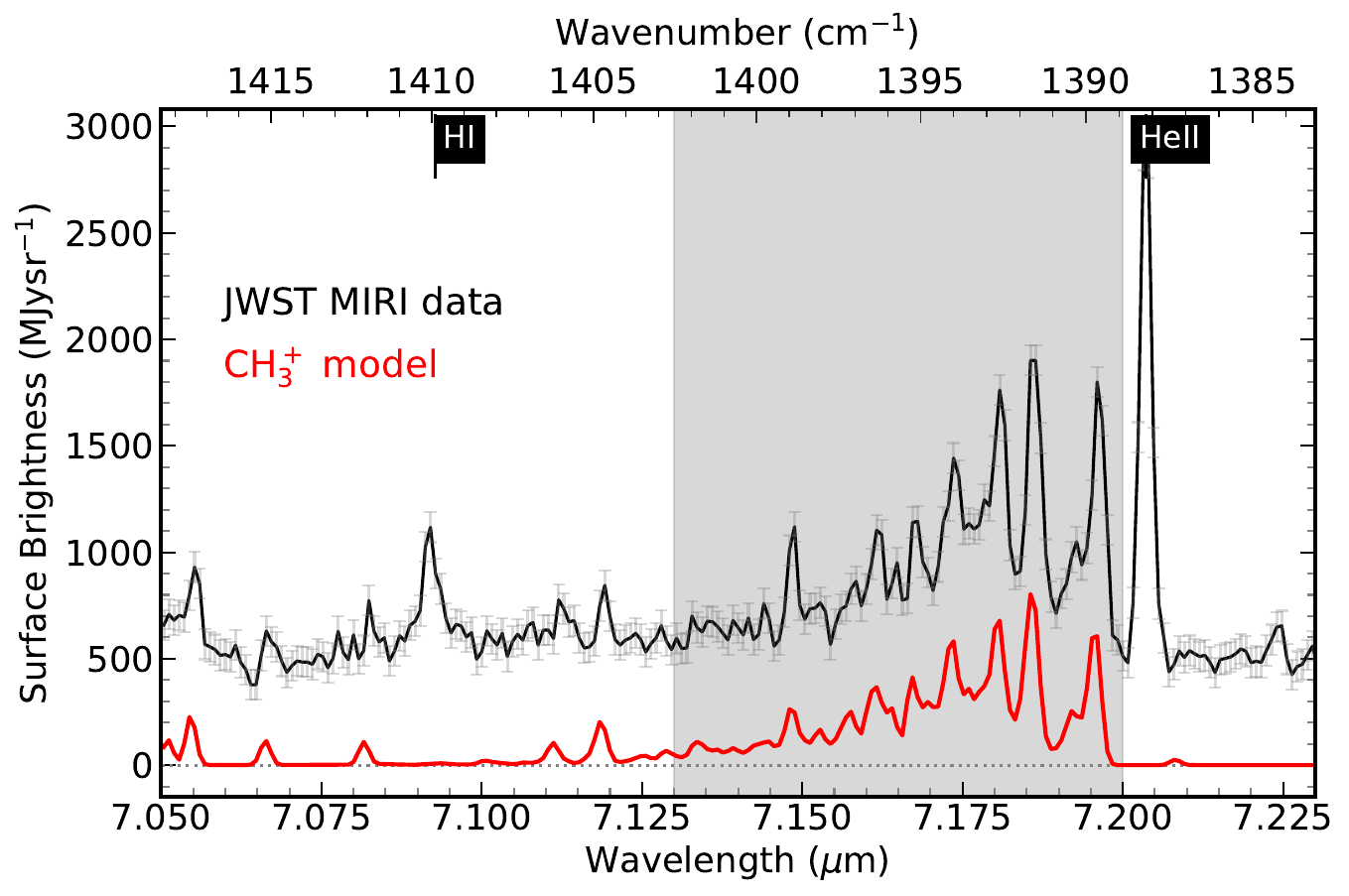}
    \caption{The continuum-subtracted JWST/MIRI spectrum of NGC 6302 at aperture 1 shown in Fig~\ref{fig:cartoon_and_apertures} (black, offset vertically for clarity) compared to a \chthreeplus model (red) by \citet{2023A&A...680A..19C} at excitation temperature of 600 K. The \ion{H}{1} recombination line at 7.0928~\micron and the \ion{He}{2} line at 7.2036~\micron are labelled at the top. }
    \label{fig:CH3+_modelvsdata}
\end{figure}

\begin{figure}
    \centering
    \includegraphics[width=\linewidth]{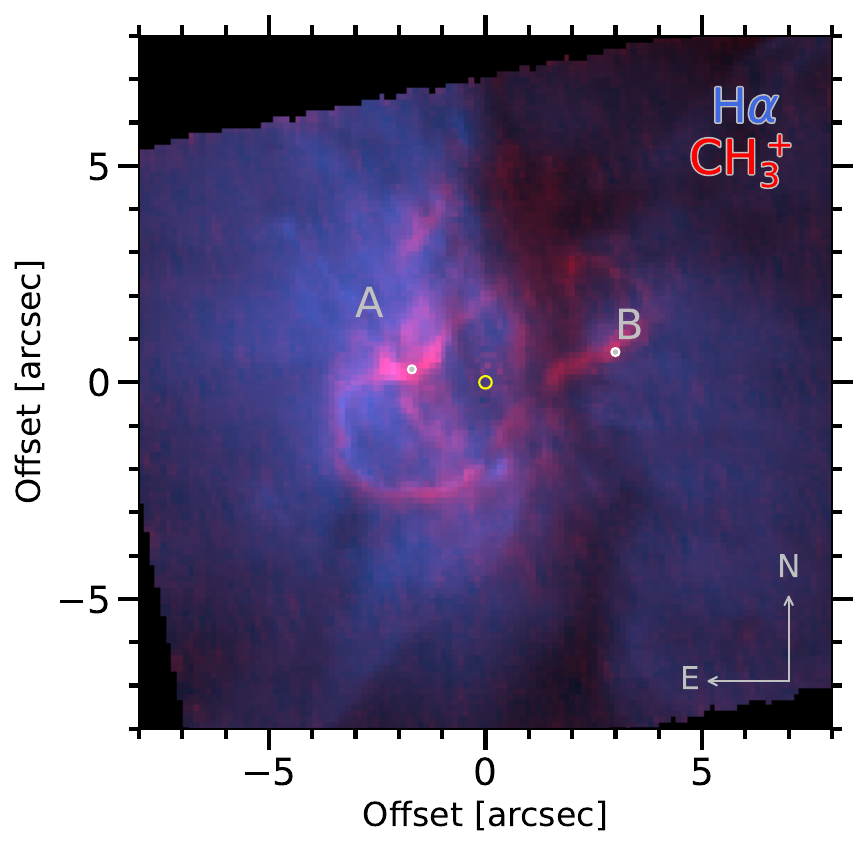}
    \caption{The image represents a colour overlay of HST/WFC3 observations featuring filter F656N in blue \citep{2022ApJ...927..100K} which traces emission from H$\alpha$, and \chthreeplus map in red.}
    \label{fig:Halpha_and_CH3+}
\end{figure}

NGC~6302 is a butterfly-shaped planetary nebula, 1.03 $\pm$ 0.27 kpc away from Earth \citep{2020MNRAS.492.4097G},  towards the constellation Scorpius. \citet{2011MNRAS.418..370W} constructed a 3D photoionization model of  NGC~6302 and constrained the properties of its central star. They derived an effective temperature of 220,000K, which explains the presence of highly ionized species such as [Mg VII] and [SiVII]. The same study yields a stellar mass of 0.73–0.82 \solar{M} (with an initial mass of about 5.5 \solar{M}), and a luminosity of 14,300~\solar{L}. NGC~6302 is O-rich with C/O $\approx$ 0.5 \citep{2011MNRAS.418..370W}. 


Optical images (see Fig.~\ref{fig:HST}) reveal two lobes of luminous gas, shaped like the wings of a butterfly. Its bipolar lobes (which stretch as far as 2.1 pc) are exceptionally bright and exhibit a rich variety of emission lines, from the far-ultraviolet to the mid-infrared \citep{2000MNRAS.314..657C, 2001ApJ...550..785F}. NGC~6302 is known for exhibiting one of the most complex morphologies seen in PNe. Optical observations (HST/WFC3) also reveal that the two lobes, which are oriented east-west, show small-scale structures such as clumps, tails, knots, and filaments, which seem to be in azimuthally organized zones. [Fe~{\sc ii}] emission reveals jets coming out from the central region in south-east and north-west directions, forming an S-shape \citep{2022ApJ...927..100K, 2023ApJ...957...54B}. 

The central region of NGC~6302 comprises three major morphological structures: the \textit{torus} (sometimes referred to as the "central dark lane" or "main ring" in previous studies), the \textit{inner bubble} (sometimes referred to as the "inner ring" in previous studies) and the \textit{outer bubble}. A schematic presentation of these morphological structures is provided in Fig.~\ref{fig:cartoon_and_apertures}. 

The torus is massive and dense with a total (dust and gas) mass of 0.8--3 \solar{M} and size of 5700 AU (\overviewpaper). It is the primary reservoir of dust in NGC~6302. These authors point out that the torus is distorted or warped at the edges and is filled with crystalline silicates and other dust grains. In optical observations, the torus appears as a central dark lane that bisects the two lobes and runs nearly north-south, orthogonal to the lobes (see Fig.~\ref{fig:HST}). That dark lane is the front part of the torus that surrounds the central star, obscuring the star in the visible and near-IR \citep{2022ApJ...927..100K}. ALMA $^{12}$CO J$=$3--2 maps reveal the torus in great detail and provide kinematic information. The torus is non-Keplerian and is radially expanding at a velocity of 8 km~s$^{-1}$. The kinematical age of the torus is estimated to be $\sim$5000--7500 years \citep{2007A&A...473..207P, 2017A&A...597A..27S}. 

The inner bubble has been observed at optical, mid-IR and sub-mm wavelengths as it is traced by emission of several species such as \ion{H}{2}, \htwo, \CO, \ce{HCO+}, PAHs, etc (\citet{2017A&A...597A..27S, 2022ApJ...927..100K}, \overviewpaper, \baezetal, \clarketal). The inner bubble, which runs east-west, is peanut-shaped and is pinched where it intersects with the previously ejected torus (\overviewpaper). The inner bubble is inclined approximately 60\textdegree~relative to the torus, and is expanding with a velocity of 11 km~s$^{-1}$. It is much younger than the torus, with a kinematical age of $\sim$2200 years \citep{2017A&A...597A..27S}. \citet{2023ApJ...957...54B} describe the sequence of several episodes of ejection from the center of the NGC~6302 in detail. 


While the torus and the inner bubble have been detected in optical and sub-mm observations before, JWST MIRI observations by \overviewpaper revealed two arcs outside the inner bubble, towards the southeast and northwest, inclined with respect to both the inner bubble and the torus (see Fig.~\ref{fig:cartoon_and_apertures}). Those arcs constitute the outer bubble, which is bright in \htwo and PAH emission. (\overviewpaper, \clarketal).

\section{\texorpdfstring{CH$_3^+$}{CH3+} emission in NGC~6302}

\label{sec:detection-of-CH3plus}

The JWST/MIRI observations of the central region of NGC~6302 reveal the characteristic \chthreeplus emission band in the 7.13--7.20$\mu$m range (see Fig.~\ref{fig:CH3+_modelvsdata}) and show that the emission is spatially extended and resolved across distinct regions.

\subsection{The spatial distribution of \texorpdfstring{\chthreeplus}{CH3+}}
\label{sec:spatial distribution of CH3+}

To reveal the spatial distribution of \chthreeplus, we integrated the surface brightness over the wavelength range 7.13--7.20~\micron after continuum subtraction for each spatial pixel in the MIRI field of view. We restricted the integration to these wavelengths to avoid contributions from other species. An example of the adopted continuum is shown in Appendix~\ref{App-sec:continuum}, and the resulting map of the integrated surface brightness is shown in Fig.~\ref{fig:cartoon_and_apertures}. The map shows bright \chthreeplus emission along the inner bubble's edge (east-west orientation) and the torus (north-south orientation). Although not visible on the map, we found that \chthreeplus is also present in the outer bubble (see below), but the column densities are too low to be visible in the integrated surface brightness map. To guide the reader and facilitate spatial comparison, we compared the \chthreeplus map to the HST/WFC3 image  \citep[through the F656N filter; see][]{2022ApJ...927..100K}, which traces H$\alpha$ emission, in Fig.~\ref{fig:Halpha_and_CH3+}. Note that the \chthreeplus emission along the torus overlaps with what appears as a central dark lane in the H$\alpha$ observations. 

To better understand what drives the \chthreeplus emission, we investigate its spectroscopic characteristics and spatial variations using defined apertures rather than individual pixels to achieve better signal-to-noise (S/N) ratios.We selected 25 apertures to sample all distinct regions with \chthreeplus emission along the torus, inner bubble, and outer bubble (see Fig.~\ref{fig:cartoon_and_apertures}). Apertures 1-22 use 2×2 pixel areas, each covering 0\farcs26×0\farcs26 given the 0$\farcs$13 pixel scale in channel 1c. Apertures 23-25 are larger elliptical regions (with areas 3\farcs22$^2$, 6\farcs52$^2$, and 4\farcs36$^2$ respectively) to achieve adequate S/N for the weaker \chthreeplus emission in the outer bubble. These apertures were positioned to cover the outer bubble arc structure as guided by the \htwo map (see Appendix~\ref{App-sec:H2map}). The aperture distribution is as follows: apertures 1-2 cover the intersection of the inner bubble and torus; apertures 3-6 and 7-11 trace the west and east edges of the inner bubble, respectively; aperture 12 lies north of the intersection point; apertures 13-15 trace the north rear side of the torus; apertures 16-20 and 21-22 trace the front and rear edges of the torus, respectively; and apertures 23-25 sample the outer bubble's northwest arc and south-southwest region.

\subsection{Spectroscopic analysis of the \texorpdfstring{\chthreeplus}{CH3+} emission}
\label{sec:T-and-N}

\begin{figure}
    \centering
    \includegraphics[width=\linewidth]{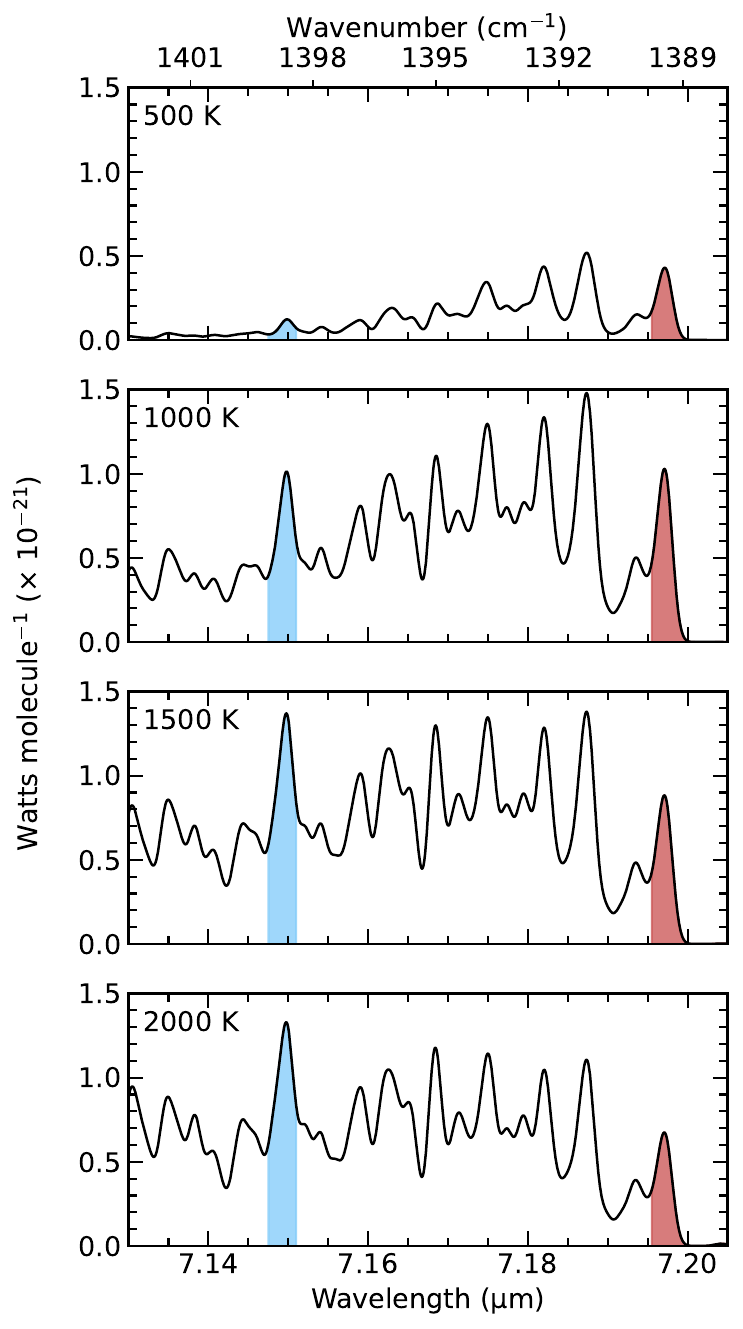}
    \caption{Model spectra of \chthreeplus at 500, 1000, 1500 and 2000 K. Bands at 7.15 and 7.2~\micron are highlighted in blue and red, respectively. }
    \label{fig:model_at_diff_temp}
\end{figure}

\begin{figure}
    \centering
    \includegraphics[width=\linewidth]{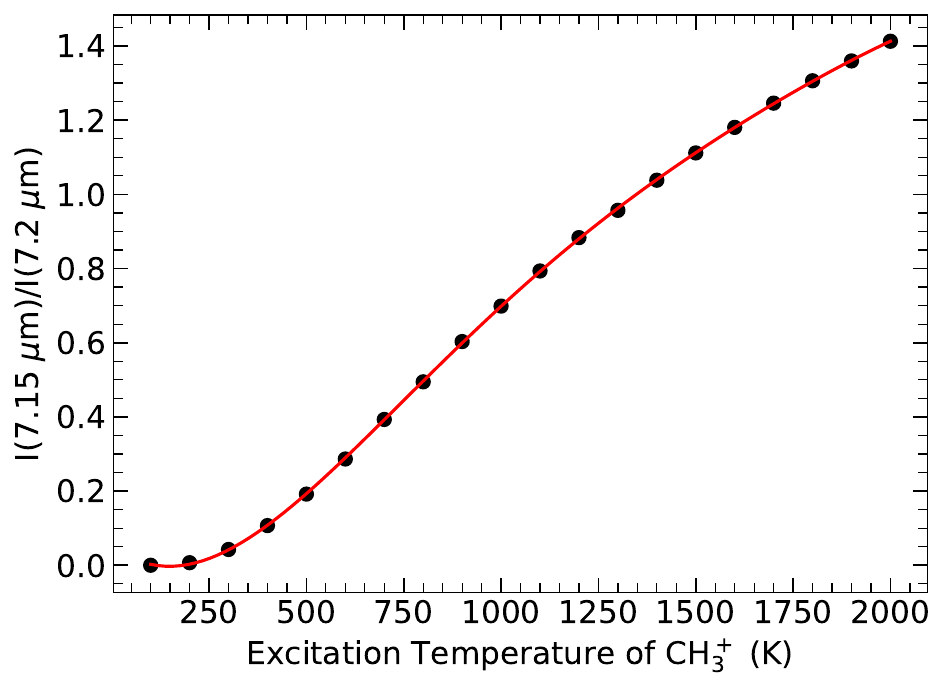}
    \caption{Intensity ratio \mbox{I(7.15~\micron)/I(7.2~\micron)} as a function of excitation temperature of \chthreeplus.  Black points represent the \mbox{I(7.15~\micron)/I(7.2~\micron)} found in the model spectra for a given excitation temperature. A five-degree polynomial fit to those points is represented by the red curve. The polynomial is given by Eq.~\ref{eq:polynomial fit}}
    \label{fig:band_ratio_vs_T}
\end{figure}

\begin{figure}
    \centering
    \includegraphics[width=\linewidth]{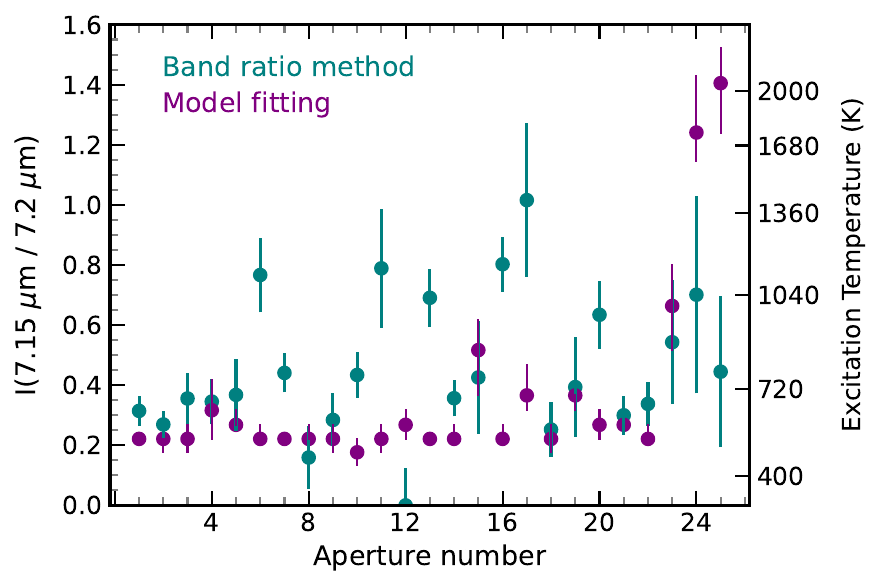}
    \caption{\chthreeplus excitation temperature derived for the 25 selected apertures. Teal points represent the \mbox{I(7.15~\micron)/I(7.2~\micron)} band found in the observed extracted spectra. The corresponding temperatures (as described  by Eq.~\ref{eq:polynomial fit}) are shown on the y-axis on the right. Purple points represent the best-fit temperature derived from fitting models to the observations.}
    \label{fig:obs_band_ratio_vs_index}
\end{figure}

\chthreeplus has a trigonal planar geometry ($D_{3h}$ point group) and it lacks a permanent dipole moment. Because of that, it cannot be observed at radio wavelengths. However, it can be detected via ro-vibrational transitions at infrared wavelengths. \chthreeplus has symmetric ($\nu_1$) and degenerate asymmetric ($\nu_3$) stretching modes at around 3~\micron ($\sim$3100 \wavenumbers), an out-of-plane umbrella bending mode ($\nu_2$) and a degenerate in-plane ($\nu_4$) bending mode at 7.2~\micron \citep[$\sim$1400 \wavenumbers; ][]{2019JChPh.150h4306M, 2023A&A...680A..19C}. Coriolis interactions between the $\nu_2$ and $\nu_4$ bands (also referred to as $\nu_2$/$\nu_4$ dyad) result in a characteristic pattern of ro-vibrational lines that occur in the 7.15--7.2~\micron range.

We compare the observations to the \chthreeplus model developed by \citet{2023A&A...680A..19C} using the latest and most detailed rovibrational assignments of the \chthreeplus feature based on theoretical and experimental studies. They performed detailed quantum mechanical calculations to derive spectroscopic constants (for the $\nu_2$/$\nu_4$ dyad) such as rotational constants, coriolis coupling constant, centrifugal distortion constants, etc., in the vibronic ground state of \chthreeplus. These constants were then validated by two different experimental methods. We use the \texttt{PGOPHER} \citep[a general-purpose software for simulating electronic, vibrational and rotational spectra; ][]{2017JQSRT.186..221W} file provided by \citep{2023A&A...680A..19C} to compute the \chthreeplus model spectrum that can be compared to the observations at 7.2~\micron at JWST's resolution. Note that the models assume a thermal distribution across rotational levels at $\nu$=1 level. We used \texttt{PGOPHER} to calculate \chthreeplus models spanning the temperature range from 100~K to 3000~K in steps of 50~K, where the spectrum is convolved the with a Gaussian adopting a full width half maximum (FWHM) of 0.37~\wavenumbers, corresponding to the JWST resolving power of R = 3600 at 7.2~\micron \citep[channel 1c; see][]{2023A&A...675A.111A}. We exported the models in the units of Watts per molecule.

Fig.~\ref{fig:model_at_diff_temp} shows how the Q-branch of the \chthreeplus model spectrum changes as a function of excitation temperature. At JWST's resolution, overlapping Q-branch lines cause a bump to appear. Note that the strength of different bands peaks at different temperatures based on the rotational levels of the transitions involved. For example, the strength of the 7.15~\micron band is lower than that of the 7.2~\micron band at lower temperatures. However, at temperatures above 1000K, the opposite is true. The intensity ratio of the 7.15~\micron band (shaded blue in Fig.~\ref{fig:model_at_diff_temp}) to the 7.20~\micron line (shaded red) is very small ($<$1) at low temperatures, $\sim$1 at about 1300~K and much higher ($>$1) at high temperatures. This band ratio is thus a sensitive diagnostic of the \chthreeplus excitation temperature. 

We will use this as a first estimate to determine the excitation temperature in the selected apertures, a diagnostic first used by \citet{2025A&A...696A..99Z} for the analysis of \chthreeplus emission in Orion Bar PDR. We fit a gaussian with a linear baseline to the 7.15~\micron (7.147 - 7.151~\micron) band and 7.20~\micron (7.195 - 7.199~\micron) band, for both models and observations. The ratio of \mbox{I(7.15~\micron)/I(7.2~\micron)} derived from the model spectra versus excitation temperature is shown in Fig.~\ref{fig:band_ratio_vs_T}, which exhibits the behaviour described above. This ratio should change gradually as a function of temperature, and to obtain a continuous form of this function, we fitted a polynomial to the data points. We found that a fifth-order polynomial fits the individual points well (see~Fig.~\ref{fig:band_ratio_vs_T}) using the following coefficients: 

\begin{equation}
\label{eq:polynomial fit}
T(x) = 1883 x^5 - 7260x^4 + 10640x^3 - 6956x^2 + 2941x + 159.6
\end{equation}

where $x$ is the band ratio. We can then use this polynomial to determine the excitation temperature using measured band ratios in the observational spectra. For each of the 25 apertures, we applied the same fitting procedure to the 7.15 µm and 7.20 µm bands as done for the models. These fits are shown in Appendix~\ref{app_sec:band-ratio}, and the intensity and uncertainties on the bands and the band ratios are listed in Table~\ref{tab:best_fit_results}. The uncertainties in the observations determine uncertainties in the intensity. The teal points in Fig.~\ref{fig:obs_band_ratio_vs_index} show the band ratio and temperatures derived for each aperture based on Eq.~\ref{eq:polynomial fit}. All apertures exhibit excitation temperatures of 500–1000 K (band ratios of 0.15–0.80). While the reported uncertainties primarily reflect random errors, systematic effects—especially those introduced by the linear baseline fit—are harder to quantify and could artificially increase or decrease the band ratios in certain apertures.

\begin{table*}[t]
  \centering
  \caption{The excitation temperature, column density and wavelength shift
for each of the 25 apertures from Fig.~\ref{fig:cartoon_and_apertures}.}
  \label{tab:best_fit_results}
  \renewcommand{\arraystretch}{1.1}
  \begin{tabular}{c|cccc|ccc}
    \hline\hline
    Aperture 
      & $I_{7.15}\,\pm\sigma$ 
      & $I_{7.20}\,\pm\sigma$ 
      & Ratio\,$\pm\sigma$ 
      & $T$\tablenotemark{a} 
      & $T$\tablenotemark{b} 
      & $N\pm\sigma_{N}$ 
      & wavelength shift \\
    & [$10^{-1}$] &[$10^{-1}$] & [$10^{-1}$]& [K] & [K] & [$10^{12}\,\mathrm{cm}^{-2}$] & [$10^{-3}\,\mu$m] \\
    \hline
    1 & $8.3\pm1.3$ &  $26.4\pm1.3$ &     $3.1\pm0.5$ &    $660_{-50 }^{+40}$ &    $550_{-0}^{+0}$ &     $13.3\pm0.2$ &     $-1.02\pm0.03$ \\    
    2 & $7.2\pm1.2$ &  $26.6\pm1.2$ &     $2.7\pm0.5$ &    $620_{-50 }^{+40}$ &    $550_{-50}^{+0}$ &     $11.4\pm0.2$ &     $-0.97\pm0.04$ \\    
    3 & $4.0\pm0.9$ &  $11.2\pm0.9$ &     $3.6\pm0.8$ &    $700_{-80 }^{+70}$ &    $550_{-50}^{+50}$ &     $3.8\pm0.2$ &     $-0.84\pm0.09$ \\    
    4 & $5.5\pm1.1$ &  $16.0\pm1.1$ &     $3.5\pm0.8$ &    $690_{-70 }^{+70}$ &    $650_{-100}^{+100}$ &     $2.0\pm0.1$ &     $-1.17\pm0.1$ \\    
    5 & $4.3\pm1.3$ &  $11.8\pm1.3$ &     $3.7\pm1.2$ &    $710_{-110 }^{+100}$ &    $600_{-0}^{+50}$ &     $5.4\pm0.1$ &     $-1.18\pm0.05$ \\    
    6 & $8.9\pm1.2$ &  $11.7\pm1.2$ &     $7.7\pm1.2$ &    $1110_{-140 }^{+150}$ &    $550_{-0}^{+50}$ &     $5.8\pm0.2$ &     $-1.19\pm0.06$ \\    
    7 & $6.7\pm0.9$ &  $15.1\pm0.9$ &     $4.4\pm0.7$ &    $770_{-60 }^{+60}$ &    $550_{-0}^{+0}$ &     $6.9\pm0.1$ &     $-1.16\pm0.04$ \\    
    8 & $1.4\pm0.9$ &  $8.6\pm0.9$ &     $1.6\pm1.0$ &    $490_{-190 }^{+120}$ &    $550_{-0}^{+50}$ &     $4.5\pm0.1$ &     $-1.19\pm0.06$ \\    
    9 & $3.5\pm1.0$ &  $12.3\pm1.0$ &     $2.8\pm0.9$ &    $630_{-100 }^{+80}$ &    $550_{-50}^{+50}$ &     $4.1\pm0.2$ &     $-1.12\pm0.1$ \\    
    10 & $7.7\pm1.2$ &  $17.7\pm1.2$ &     $4.3\pm0.8$ &    $770_{-70 }^{+70}$ &    $500_{-50}^{+50}$ &     $5.8\pm0.2$ &     $-1.12\pm0.06$ \\    
    11 & $4.8\pm1.0$ &  $6.1\pm1.0$ &     $7.9\pm2.0$ &    $1140_{-220 }^{+250}$ &    $550_{-50}^{+50}$ &     $3.3\pm0.2$ &     $-0.9\pm0.1$ \\    
    12 & $0.0\pm1.7$ &  $13.7\pm1.7$ &     $0.0\pm1.2$ &    $160_{-490 }^{+270}$ &    $600_{-50}^{+50}$ &     $5.7\pm0.1$ &     $-1.01\pm0.05$ \\    
    13 & $8.0\pm0.9$ &  $11.6\pm0.9$ &     $6.9\pm1.0$ &    $1020_{-110 }^{+110}$ &    $550_{-0}^{+0}$ &     $5.4\pm0.1$ &     $-0.95\pm0.05$ \\    
    14 & $3.7\pm0.6$ &  $10.4\pm0.6$ &     $3.6\pm0.6$ &    $700_{-50 }^{+50}$ &    $550_{-0}^{+50}$ &     $4.5\pm0.1$ &     $-1.1\pm0.07$ \\    
    15 & $2.3\pm0.9$ &  $5.4\pm0.9$ &     $4.3\pm1.9$ &    $760_{-170 }^{+180}$ &    $850_{-150}^{+100}$ &     $1.0\pm0.1$ &     $-0.74\pm0.13$ \\    
    16 & $10.1\pm0.9$ &  $12.5\pm0.9$ &     $8.0\pm0.9$ &    $1150_{-110 }^{+120}$ &    $550_{-0}^{+50}$ &     $4.6\pm0.2$ &     $-1.14\pm0.07$ \\    
    17 & $4.2\pm0.7$ &  $4.1\pm0.7$ &     $10.2\pm2.6$ &    $1430_{-320 }^{+380}$ &    $700_{-50}^{+100}$ &     $1.4\pm0.1$ &     $-1.13\pm0.12$ \\    
    18 & $3.4\pm1.2$ &  $13.6\pm1.2$ &     $2.5\pm0.9$ &    $600_{-110 }^{+90}$ &    $550_{-50}^{+50}$ &     $4.7\pm0.2$ &     $-1.25\pm0.07$ \\    
    19 & $1.7\pm0.7$ &  $4.2\pm0.7$ &     $3.9\pm1.7$ &    $730_{-160 }^{+150}$ &    $700_{-50}^{+0}$ &     $1.8\pm0.1$ &     $-1.34\pm0.09$ \\    
    20 & $5.4\pm0.8$ &  $8.6\pm0.8$ &     $6.3\pm1.1$ &    $960_{-110 }^{+130}$ &    $600_{-50}^{+50}$ &     $2.2\pm0.1$ &     $-1.16\pm0.08$ \\    
    21 & $3.2\pm0.7$ &  $10.8\pm0.7$ &     $3.0\pm0.7$ &    $650_{-70 }^{+60}$ &    $600_{-0}^{+0}$ &     $4.8\pm0.1$ &     $-0.81\pm0.05$ \\    
    22 & $4.6\pm0.9$ &  $13.5\pm0.9$ &     $3.4\pm0.7$ &    $680_{-70 }^{+60}$ &    $550_{-0}^{+50}$ &     $4.8\pm0.1$ &     $-0.85\pm0.05$ \\    
    23 & $0.7\pm0.2$ &  $1.4\pm0.2$ &     $5.4\pm2.1$ &    $870_{-180 }^{+220}$ &    $1000_{-150}^{+150}$ &     $0.2\pm0.0$ &     $-0.96\pm0.11$ \\    
    24 & $1.0\pm0.4$ &  $1.4\pm0.4$ &     $7.0\pm3.3$ &    $1030_{-320 }^{+410}$ &    $1750_{-150}^{+350}$ &     $0.4\pm0.0$ &     $-1.05\pm0.1$ \\    
    25 & $0.9\pm0.5$ &  $2.0\pm0.5$ &     $4.4\pm2.5$ &    $780_{-240 }^{+250}$ &    $2050_{-300}^{+500}$ &     $0.4\pm0.0$ &     $-1.05\pm0.11$ \\    
    \hline
  \end{tabular}
\tablenotetext{a}{Band ratio method}
\tablenotetext{b}{Model fitting}
\end{table*}

We can also determine the excitation temperatures by directly fitting model spectra to the continuum-subtracted spectra, with the temperature and column density of \chthreeplus as free parameters. We used the continuum-subtracted flux (in  \(\text{W m}^{-2} \, \mu\text{m}^{-1}\)) in the range of 7.13--7.199~\micron to fit models to observations. We used the \texttt{lmfit} \citep{2014zndo.....11813N} python package, adopting a Levenberg-Marquardt algorithm to determine temperatures and column densities; the excitation temperatures, however, were constrained on a grid with values between 100~K and 3000~K in steps of 50~K. We also allowed for small wavelength shifts ($\sim$ $\pm$0.0015~\micron) during fitting to account for radial velocity corrections. Appendix~\ref{sec:fit_results} shows the resulting best fits for all 25 apertures, and the resulting best fit parameters are listed in Table~\ref{tab:best_fit_results}. Parameter uncertainties presented in Table~\ref{tab:best_fit_results} denote 1$\sigma$ confidence intervals, obtained from the projections of the $\Delta\chi^{2}=1$ contour on the $\chi^{2}$ surface. The models generally reproduce the observations extremely well, leaving only very small residuals in this wavelength range. We find that the excitation temperature is in the range of 500--800K for apertures 1-22 and temperature 1000-2000K for apertures 23-25; these are represented by purple points in Fig~\ref{fig:obs_band_ratio_vs_index}.

Although no clear patterns exist in excitation temperature variations across the inner bubble and torus, the outer bubble exhibits significantly higher temperatures (1000–2000 K) compared to 500–800 K elsewhere.  The column densities (Table~\ref{tab:best_fit_results}) are typically a few times $10^{12}$ cm$^{-2}$, ranging about two orders of magnitude from the lowest (aperture 23) to highest (aperture 1) values. The highest column density is seen where the inner bubble runs into the rear side of the torus. These column densities are however, not an accurate representation of the total \chthreeplus column density: as was pointed out by \citet{2023Natur.621...56B} and \citet{2025A&A...696A..99Z} since the \chthreeplus emission originates from chemical pumping and consequently, the excitation of the vibrational levels is far from LTE.


\section{Discussion}
\label{sec:discussion}


\begin{figure*}[t]
    \gridline{
    \fig{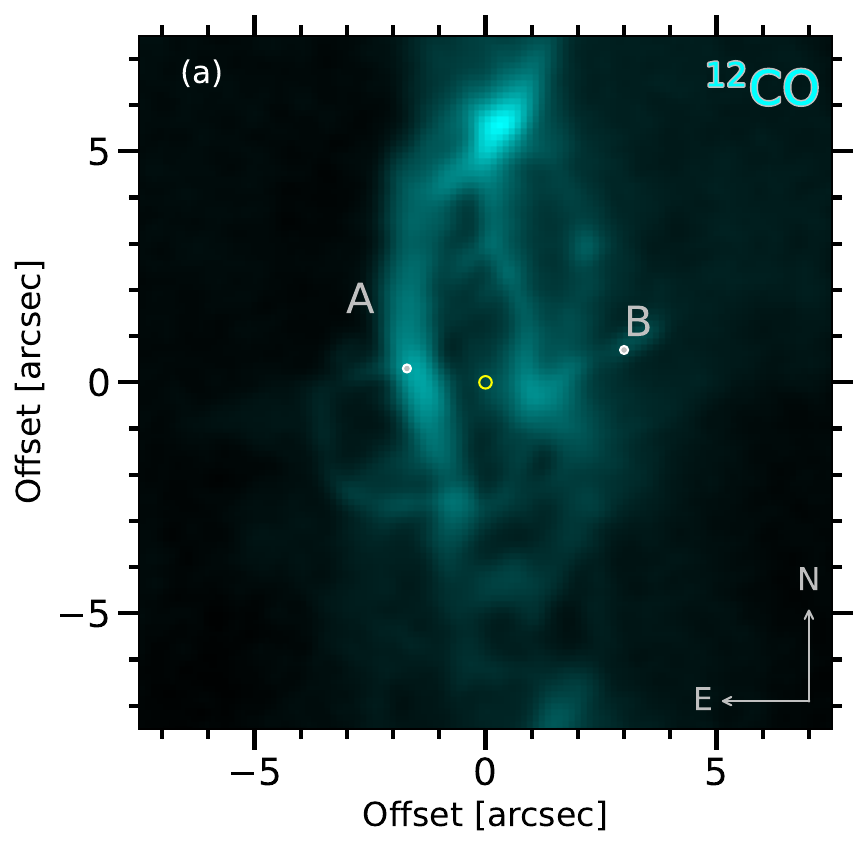}{0.45\textwidth}{}
    \fig{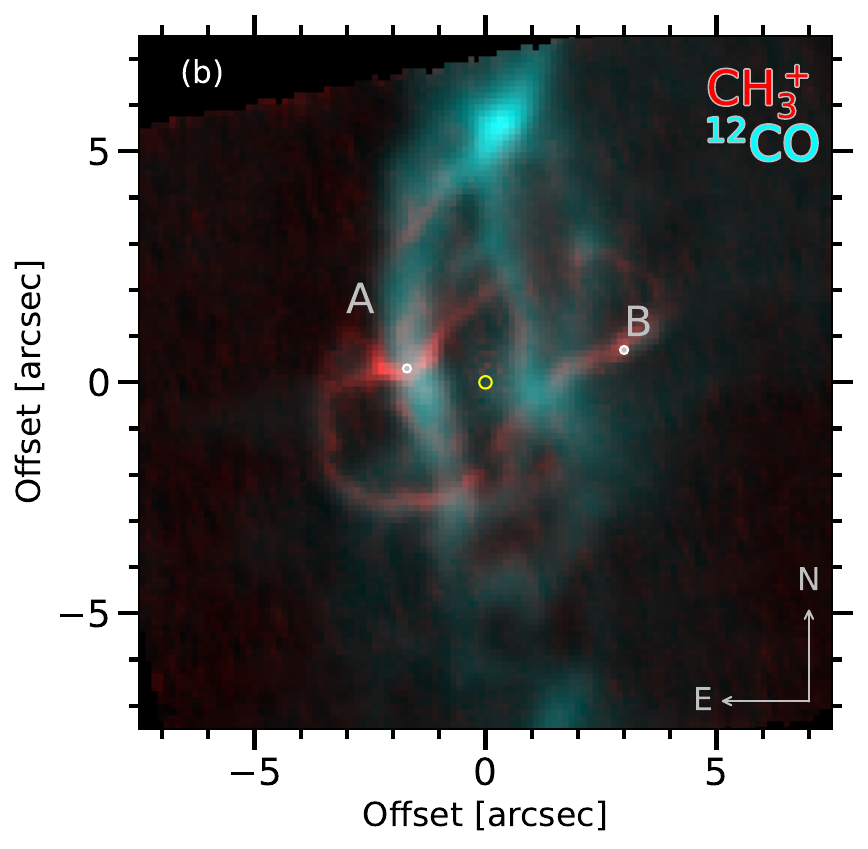}{0.45\textwidth}{}
    }
    \gridline{
    \fig{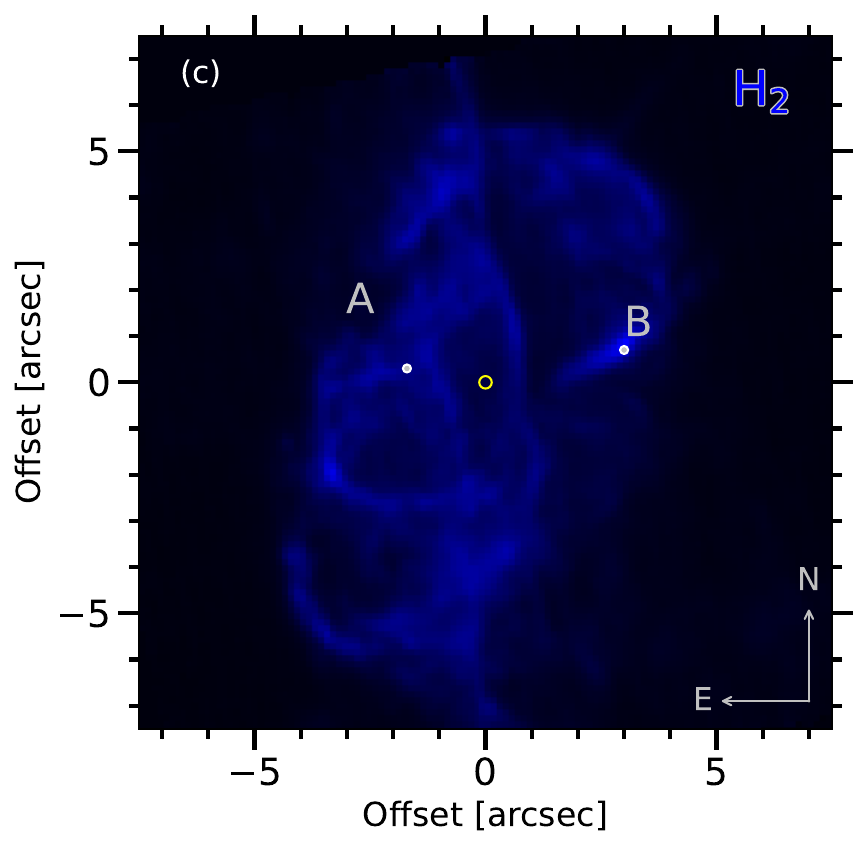}{0.45\textwidth}{}
    \fig{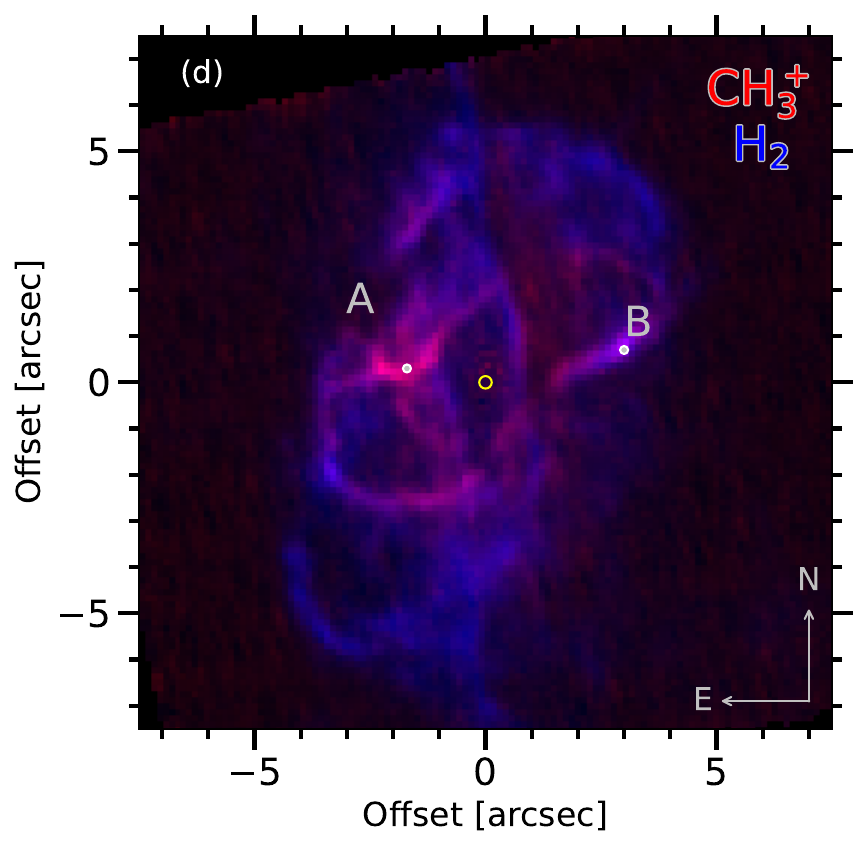}{0.45\textwidth}{}
    }
  \caption{Comparision of the integrated surface brightness maps of \chthreeplus, \htwo and \CO. \textbf{(a)}\CO(2--1, 230GHz, by \baezetal). \textbf{(b) }Two colour composite image: \chthreeplus (red) and \CO (cyan). \textbf{(c)} \htwo 0-0 S5 at 6.908~\micron \textbf{(d)} Two colour composite image: \chthreeplus (red) and \CO (cyan). Locations marked by A and B represent the two bright spots on \chthreeplus map. The maps are centred at the central star (R.A.= 17$^h$:13$^m$;44.488 $\pm$ 0.004$^s$, Dec. = -37$^\circ$:06$'$:11.76 $\pm$ 0.03$'$) whose position is indicated by the yellow circle at the center.  }
  \label{fig:CH3plus_formation_6plots}
\end{figure*}

\begin{figure*}[t]
    \centering
    \includegraphics[width=\textwidth]{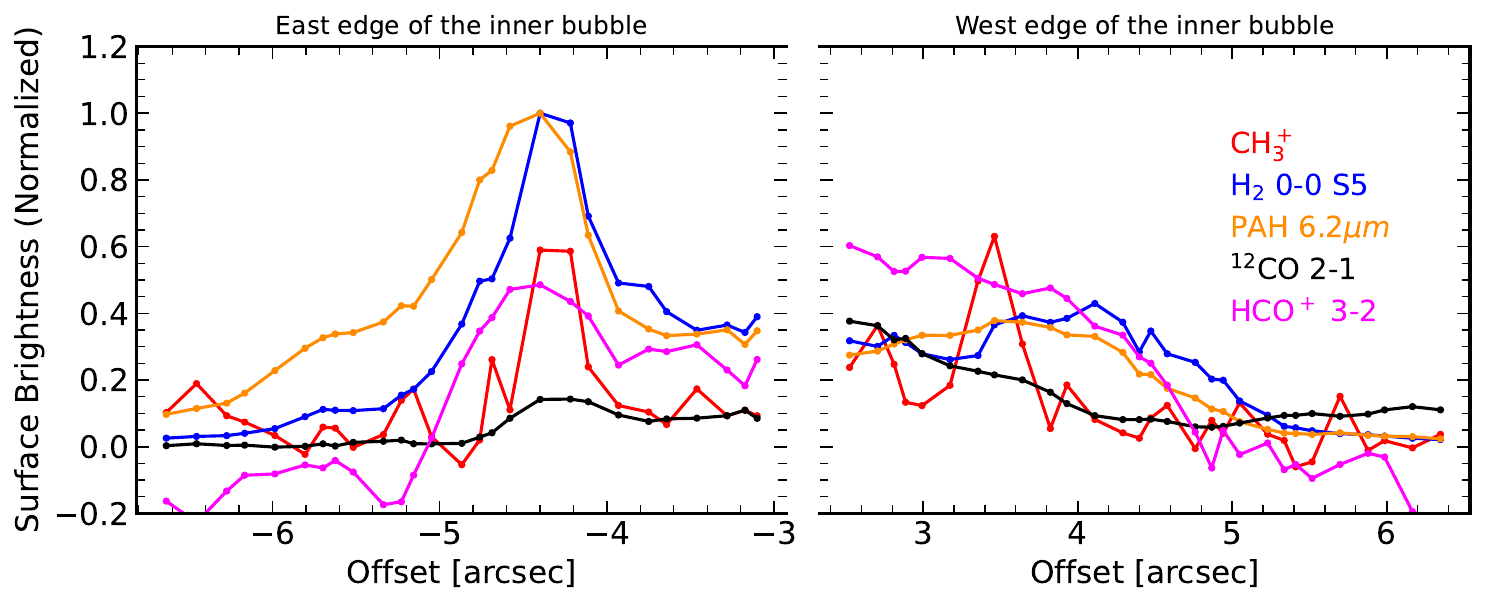}
    \caption{\textit Normalized surface brightness profiles of \chthreeplus, \htwo, PAH 6.2~\micron (by \clarketal), CO and HCO$^+$ (by \baezetal) along the cut. Offset is measured in arcsec from the position of the central star (R.A.= 17$^h$:13$^m$;44.488 $\pm$ 0.004$^s$, Dec. = -37$^\circ$:06$'$:11.76 $\pm$ 0.03$'$).}
    \label{fig:crosscuts}
\end{figure*}

\subsection{Formation of \texorpdfstring{\chthreeplus}{CH3+} in NGC~6302}
\label{sec:discussion-formation_of_CH3plus}

The detection of \chthreeplus in NGC~6302 was surprising at first because this is an O-rich PN where most carbon is locked up in CO, making carbon chemistry unlikely. However, intense UV radiation from the central star fundamentally changes this scenario by providing two key ingredients required for the formation of \chthreeplus: C$^+$  ions and vibrationally excited \htwo. 
C$^+$ reacts with the vibrationally excited \htwo to form \chplus, followed by two successive exothermic hydrogen abstraction reactions to produce \chthreeplus \citep{2023Natur.621...56B,2025A&A...696A..99Z}. The UV radiation plays a crucial role because it dissociates and photoionizes CO, thereby providing C$^+$. Moreover, in such environments, the energy barrier of the \chplus step ($\approx$0.40\,eV or $\sim$4640\,K ) is overcome by FUV-pumped \htwo, which leaves \htwo in a vibrationally excited state \citep{2010ApJ...713..662A,2013A&A...550A..96N,2013A&A...550A...8G}. \chplus and \chthreeplus form in a vibrationally excited state and then relax radiatively. This excitation mechanism is called chemical (formation) pumping, and the observed emission thus traces the newly formed \chplus and \chthreeplus. 

Our observations confirm this formation pathway in NGC~6302 through spatial comparison of \CO, \htwo and \chthreeplus. Fig.~\ref{fig:CH3plus_formation_6plots} shows that \CO and \htwo are indeed present where \chthreeplus emission is detected i.e. where \chthreeplus is formed. \chthreeplus and \CO are also present in the outer bubble, but as mentioned above, their column densities are too low to be visible in the integrated surface brightness map. Since the vibrationally excited \htwo lines detected with JWST/MIRI data have very low S/N, their spatial distribution has large uncertainties for comparison with \chthreeplus distribution. We therefore use the H$_2$ 0-0 S(5) line (at 6.908~\micron) as a tracer for the \htwo distribution.

Fig.~\ref{fig:crosscuts} shows the normalized surface brightness profile of these species along the cut indicated in Fig.~\ref{fig:cartoon_and_apertures}, zoomed in to show the east and west edges of the bubble. On the east edge, \chthreeplus, \CO and \htwo show peaks in the surface brightness profiles at the same distance from the star, with \chthreeplus and \htwo having very similar profiles. The west edge shows a very different distribution, likely because of projection effects. Notably, \chthreeplus exhibits the sharpest rise and fall at the inner bubble edge compared to other species, including \htwo. This could indicate where FUV-pumped \htwo (v$>$0) is available; however, NIRSpec observations of vibrationally excited \htwo lines are needed for confirmation. Alternatively, this may reflect the influence of fast winds or energetic shocks that formed the inner bubble (\overviewpaper), confining \chthreeplus formation to a narrow region along the edge.

\subsection{Spatial variations in \texorpdfstring{CH$_3^+$}{CH3+} emission}

\chthreeplus emission shows significant spatial variations in column density across NGC 6302. The most prominent features are two bright spots, Points A and B, which exhibit similar temperatures to other regions but significantly higher column densities (See Fig.~\ref{fig:cartoon_and_apertures}). This enhancement occurs where the inner bubble runs into the torus, giving the inner bubble its characteristic peanut shape and creating regions of higher gas density. Point A (region around apertures 1,2, and 12), exhibiting the brightest \chthreeplus emission, is located at the intersection of the inner bubble with the rear side of the torus. A corresponding bright feature is not observed where the inner bubble intersects the front side because of the extinction by the dust in the torus. However, enhanced emission is detected in both the east (aperture 7) and west (apertures 5 and 6) of this intersection, potentially tracing redirected material emerging from the intersection. This spot on the west is designated point B. Apart from higher gas density, secondary factors may also contribute to enhanced formation rates. For example, at point A, the \ion{H}{1} emission is several times brighter than the surrounding regions, suggesting higher UV flux, and point B coincides with a brighter \htwo emission that could boost \chthreeplus production. 

In contrast, the outer bubble shows column densities an order of magnitude lower than typical values in the inner bubble and the torus and two orders of magnitude lower than point A. This is likely due to the lower column density of CO in the outer bubble. Interestingly, the \chthreeplus excitation temperature is significantly higher in the outer bubble (1000-2000K) than what is found in the inner bubble and torus (500-800K). The chemical pumping model by \citet{2025A&A...696A..99Z} shows that higher excitation temperature of \chplus results from enhanced \htwo level population densities, which occur when \htwo is more efficiently FUV-pumped to higher vibrational states. Similarly, hotter \chthreeplus can form if the \htwo involved in hydrogen abstraction reactions is vibrationally excited to higher levels. This indicates that the outer bubble is less shielded than the inner structures.

\subsection{Chemistry initiated by \texorpdfstring{\chthreeplus}{CH3+}}

\begin{figure*}[t]
    \resizebox{\hsize}{!}{
    \includegraphics{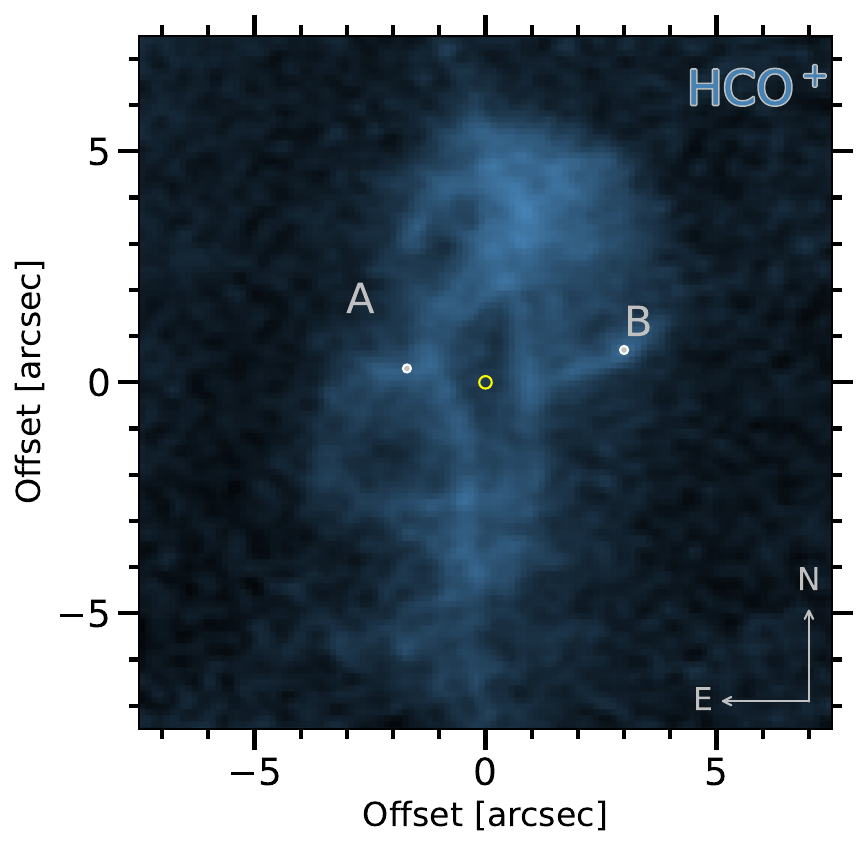}
    \includegraphics{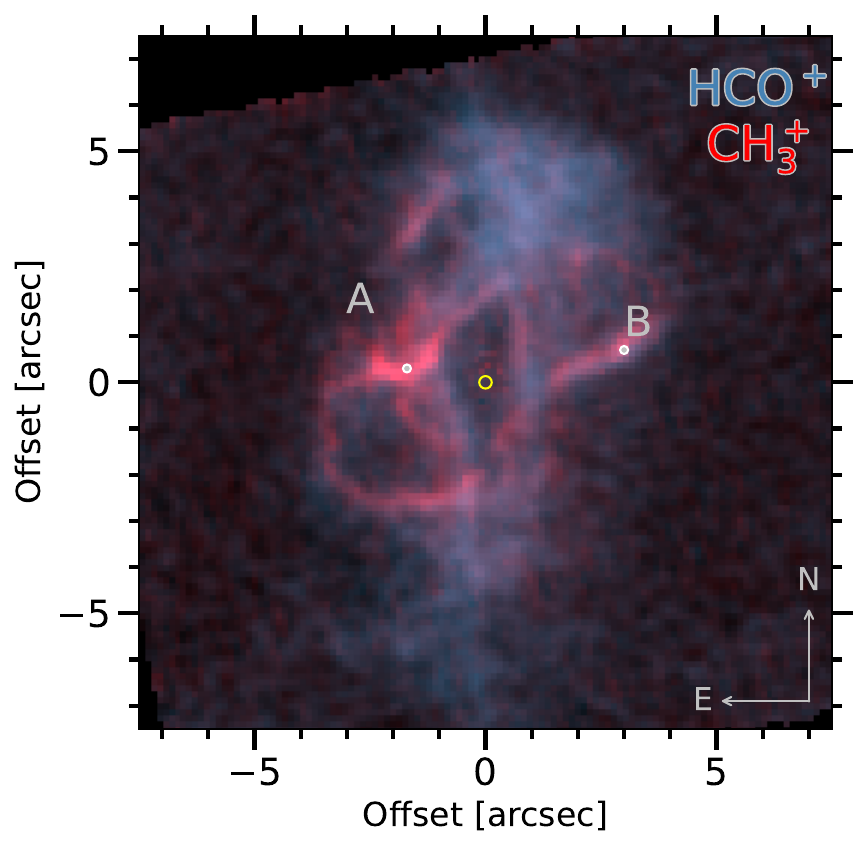}
    }
    \caption{HCO$^+$ map (by \baezetal) is shown in blue on the left and an overlay of \chthreeplus map is shown in red on the right. Locations marked by A and B represent the two bright spots on \chthreeplus map. The maps are centred at the central star (RA=17:13:44.488±0.004, dec=-37:06:11.76±0.03) whose position is indicated by the white circle at the center.   }
    \label{fig:HCOp_and_CH3plus}
\end{figure*}

\begin{figure*}[t!]
\resizebox{\hsize}{!}{\includegraphics{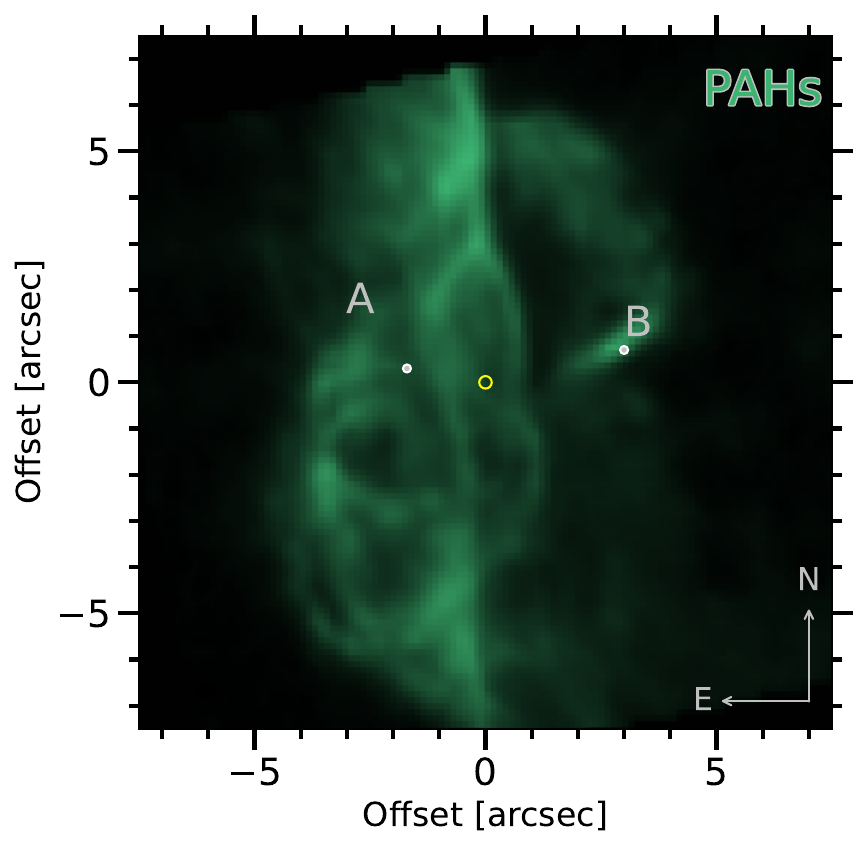}\includegraphics{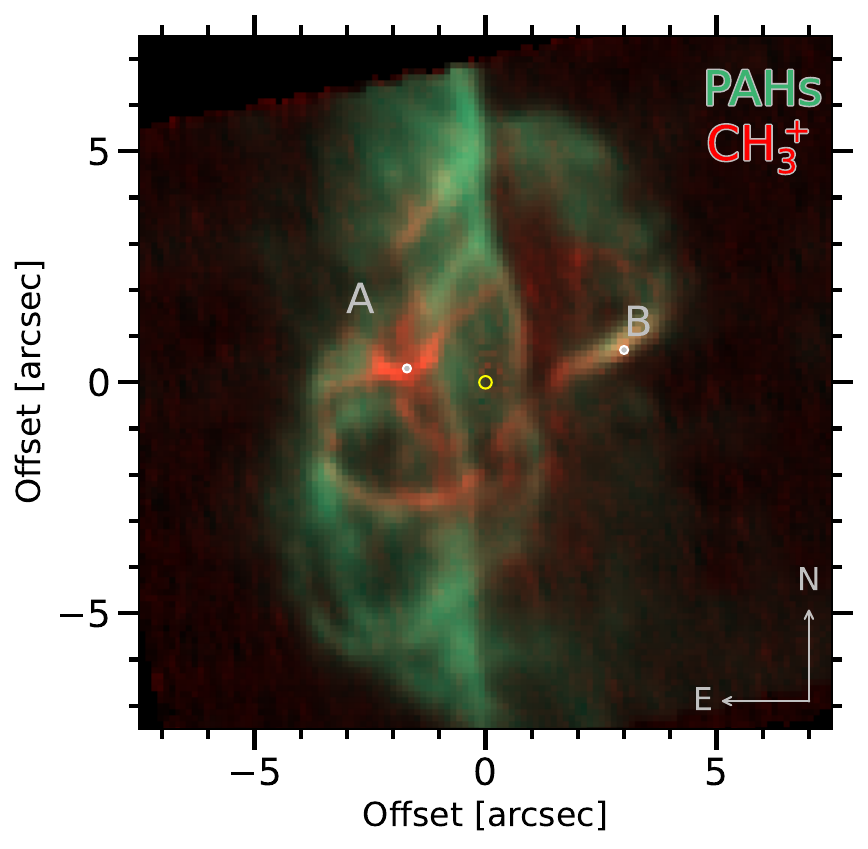}}
     \caption{The PAH map (by \clarketal) here on the left shows the integrated intensity of the 6.2~\micron feature (integrated over the 6.12--6.5~\micron range after continuum subtraction). A comparison with \chthreeplus map is shown on the right. Locations marked by A and B represent the two bright spots on \chthreeplus map. The maps are centred at the central star (R.A.= 17$^h$:13$^m$;44.488 $\pm$ 0.004$^s$, Dec. = -37$^\circ$:06$'$:11.76 $\pm$ 0.03$'$) whose position is indicated by the white circle at the center.   }
    \label{fig:PAH_and_CH3plus}
\end{figure*}

The main takeaway from the \chthreeplus detection in NGC~6302 is recognizing that hydrocarbon radical chemistry, previously overlooked in oxygen-rich environments, should be incorporated into planetary nebula chemical models. \chthreeplus drives gas-phase carbon chemistry through exothermic ion-neutral reactions, with models showing it can facilitate the formation of numerous species, including CN, HCN, CH$_2$CO$^+$, CH$_3$OH, H$_2$CO, C$_2$H$_5$OH, CH$_3$OCH$_3$, H$_2$CN$^+$, etc. (see Fig. 2 by \citet{smith_ion_1992} for the summary of the chemical network initiated by \chthreeplus). 

Spatial coincidences in NGC~6302 suggest potential connections between \chthreeplus and other observed species that warrant consideration in chemical models. \chthreeplus is spatially coincident with \ce{HCO+} along the inner and outer bubble edges (see Fig.~\ref{fig:HCOp_and_CH3plus}), consistent with the exothermic reaction $\mathrm{CH_3^+ + O \rightarrow HCO^+ + H_2}$ \citep{2000JChPh.112.4959S}. The co-location of \chthreeplus with PAHs (see Fig.~\ref{fig:PAH_and_CH3plus}) raises the question whether \chplus or \chthreeplus might initiate chemistry relevant to PAH formation. \citet{2008A&A...483..831A} reported that strong FUV/X-ray radiation can efficiently drive carbon chemistry in dense O-rich gas where \chplus and \chthreeplus can further react with e$^-$, C, and \htwo to form C$_2$H$_2$, which can then initiate the chemistry leading to aromatic species like PAHs if activation energies are overcome. Since PAH emission is widespread in NGC~6302's central region—including the torus, inner and outer bubbles where \chthreeplus is detected—the role of \chthreeplus in the bottom-up PAH formation sequence should be investigated.

\section{Summary \& Conclusions}
\label{sec:conclusion}

We report the detection of methyl cation (\chthreeplus) in JWST-MIRI observations of the O-rich planetary nebula NGC 6302, marking the first identification of this key organic precursor in such an environment. We investigated the nature of the \chthreeplus emission and its relation to other species such as \CO, \htwo, PAHs and HCO$^+$. We fit the observed \chthreeplus emission with LTE models to derive excitation temperature and column density. Model fits reproduce the \chthreeplus emission very well, yielding excitation temperatures of 500-800K in the inner bubble and the torus, significantly higher excitation temperature of 1000-2000K in the outer bubble. Column densities of \chthreeplus in the excited state are typically $\sim$10$^{12}$ cm$^{-2}$ ranging from $\sim$10$^{11}$ cm$^{-2}$ in the outer bubble to $\sim$10$^{13}$ cm$^{-2}$ where the inner bubble runs into the torus. These measurements represent only a fraction of the total population due to non-LTE chemical pumping. The spatial coincidence of \chthreeplus with \CO and \htwo supports its formation via chemical pumping in the UV-irradiated O-rich gas.

The presence of \chthreeplus in this O-rich PN highlights the carbon chemistry at play in O-rich environments under strong UV irradiation. These results indicate that hydrocarbon radical chemistry must be incorporated into chemical models of PNe. Further observations at comparable spatial resolution, particularly in the near-IR (which is rich in molecular features), are essential to unravel the different chemical networks involved in this chemistry.

\begin{acknowledgments}

This work is based on observations made with the NASA/ESA/CSA James Webb Space Telescope. All of the data presented in this article were obtained from the Mikulski Archive for Space Telescopes (MAST) 
at the Space Telescope Science Institute. The data of this specific observing program can be accessed via \dataset[doi:10.17909/s1rn-1t84]{http://archive.stsci.edu/doi/resolve/resolve.html?doi=10.17909/s1rn-1t84}. This study is based on the international consortium of ESSENcE (Evolved Stars and their Nebulae in the JWST era). This article/publication is based upon work from COST Action NanoSpace, CA21126, supported by COST (European Cooperation in Science and Technology).

C.B, J.C., E.P. and N.C. acknowledge support from the University of Western Ontario, the Canadian Space Agency (CSA)[22JWGO1-22], and the Natural Sciences and Engineering Research Council of Canada. 
M.M. and R.W. acknowledge support from the STFC Consolidated grant (ST/W000830/1).
M.J.B. and R.W. acknowledge support from the European Research Council (ERC) Advanced Grant SNDUST 694520. 
A.A.Z. acknowledges funding through UKRI/STFC through grant ST/T000414/1.
H.L.D. acknowledges support from grant JWST-GO-01742.004 and NSF grants 1715332 and 2307117.
N.C.S.\ acknowledges support from NSF award AST-2307116.
G. G.-S. thanks Michael L. Norman and the Laboratory for Computational Astrophysics for the use of ZEUS-3D. The computations were performed at the Instituto de Astronom{\'i}a-UNAM at Ensenada.
P.J.K. acknowledges support from the Science Foundation Ireland/Irish Research Council Pathway programme under Grant Number 21/PATH-S/9360.
R.S.’s contribution to the research described here was carried out at the Jet Propulsion Laboratory, California Institute of Technology, under a contract with NASA.
K.E.K. acknowledges support from grant JWST-GO-01742.010-A.
F.K. and M.T. acknowledge support from the Spanish Ministry of Science, Innovation and Universities, under grant number PID2023-149918NB-I00. This research was also partly supported by the Spanish program Unidad de Excelencia María de Maeztu CEX2020-001058-M, financed by MCIN/AEI/10.13039/501100011033.
 JML was supported by basic research funds of the Office of Naval Research.

\end{acknowledgments}

\appendix

\section{Continuum}

\begin{figure}
    \centering
    \includegraphics[width=\linewidth]{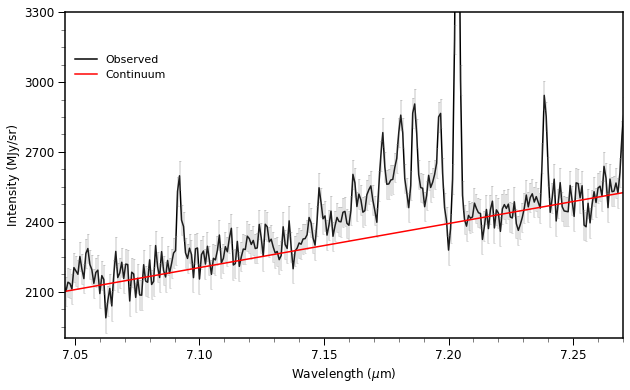}
    \caption{JWST MIRI spectrum at aperture 1 (in black). The red line provides an example of the continuum adopted to fit the models to the observations. }
    \label{fig:continuum_CH3plus}
\end{figure}

An example of the continuum adopted for fitting the \chthreeplus spectral features is shown in Fig.~\ref{fig:continuum_CH3plus}. 
\label{App-sec:continuum}

\section{\texorpdfstring{H$_2$}{H2} map with apertures}

\begin{figure}
    \centering
    \includegraphics[width=\linewidth]{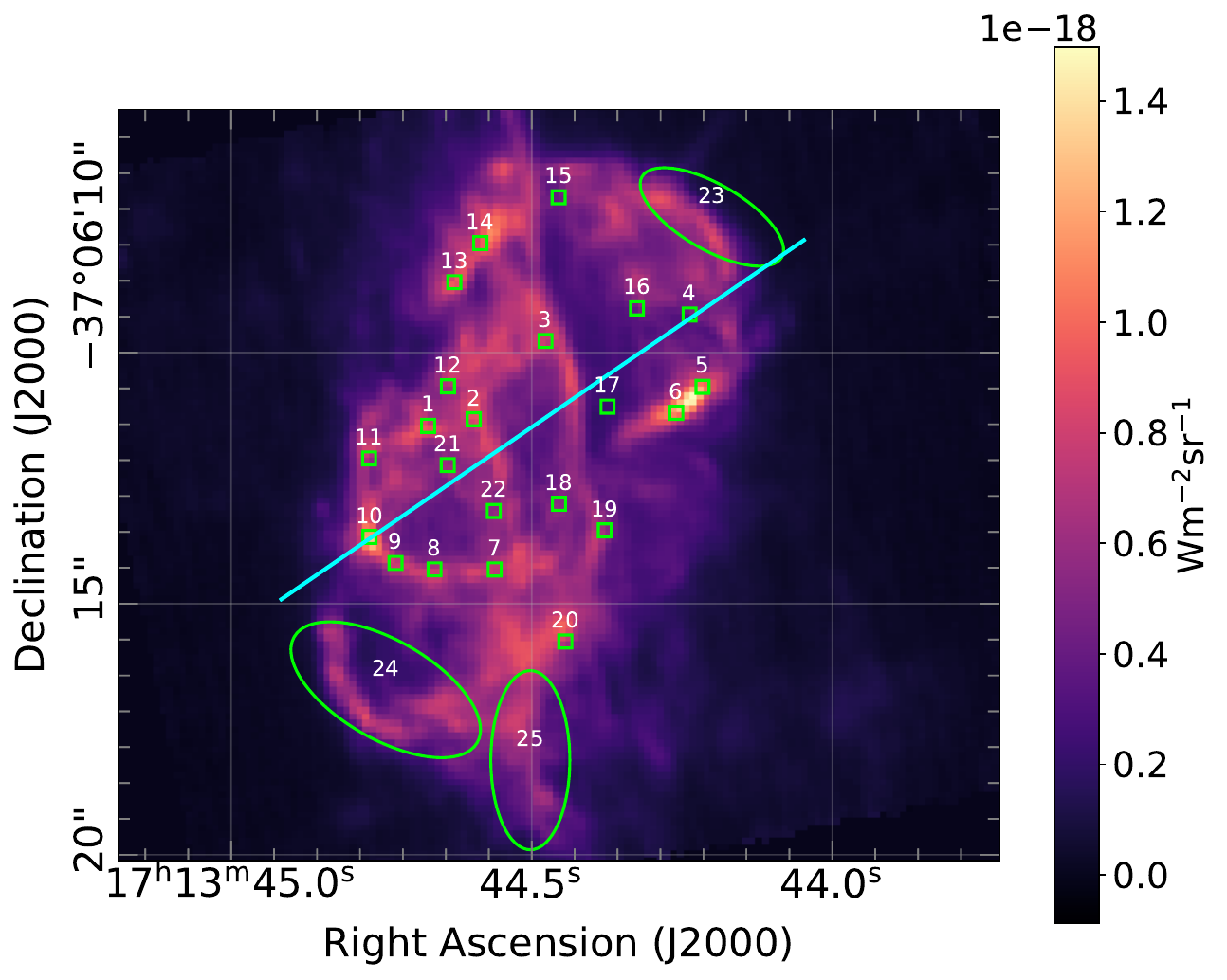}
    \caption{Integrated surface brightness map of \htwo 0-0 S(5) line at 6.908~\micron, The green boxes and ellipses represent the 25 apertures used to study \chthreeplus emission in detail.  }
    \label{fig:aper_onH2map}
\end{figure}

The 25 apertures used to sample the \chthreeplus emission across distinct regions are plotted on the \htwo map in Fig.~\ref{fig:aper_onH2map}. 
\label{App-sec:H2map}

\section{Band ratio method}

\begin{figure*}
    \centering
    \includegraphics[width=\textwidth]{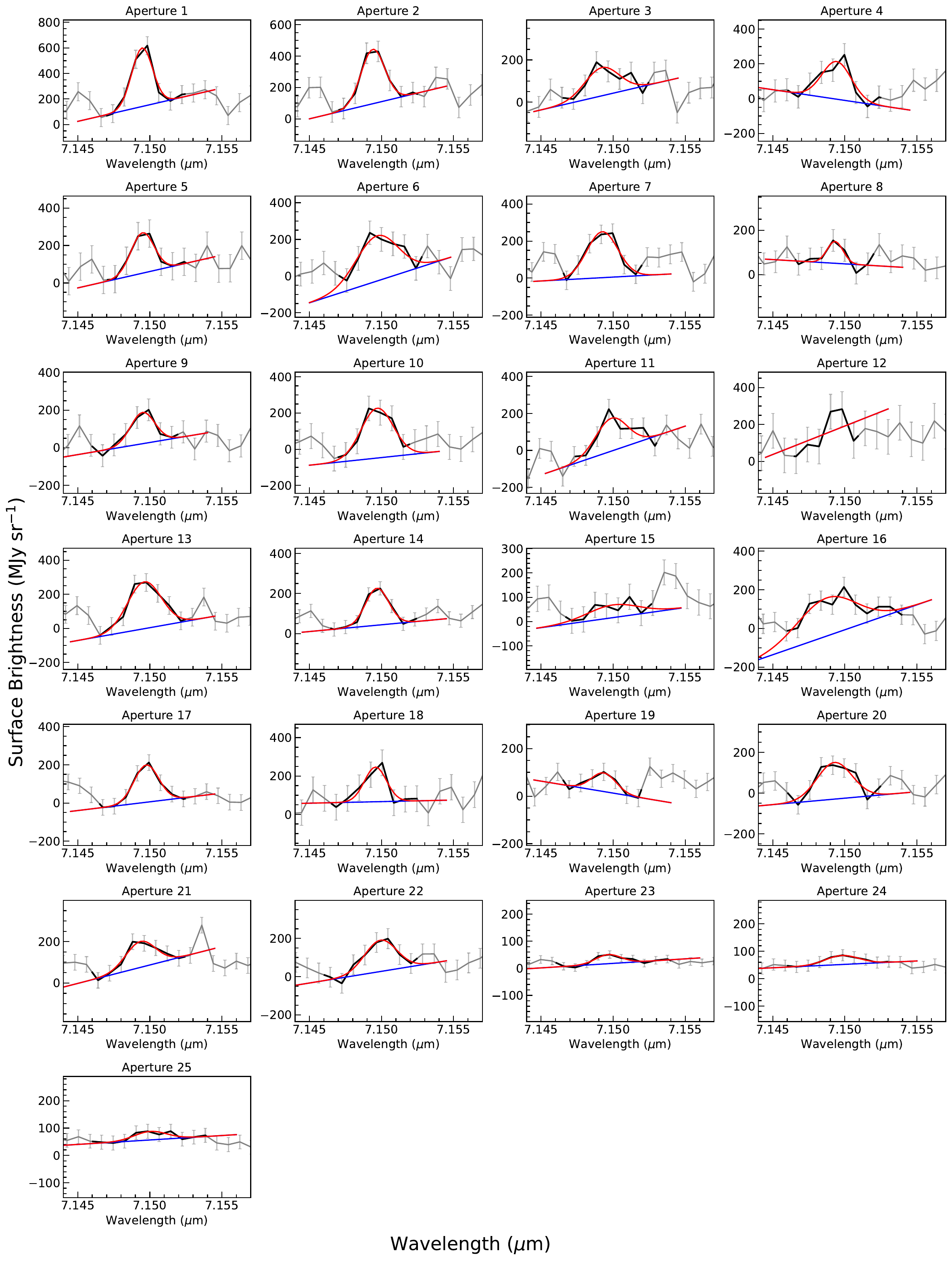}
    \caption{Gaussian fits to the 7.15 band used for the band ratio method to determine the temperature. The integrated surface brightness value at each aperture is listed in Table~\ref{tab:best_fit_results}. }
    \label{fig:gaussfits_7p15}
\end{figure*}

\begin{figure*}
    \centering
    \includegraphics[width=\textwidth]{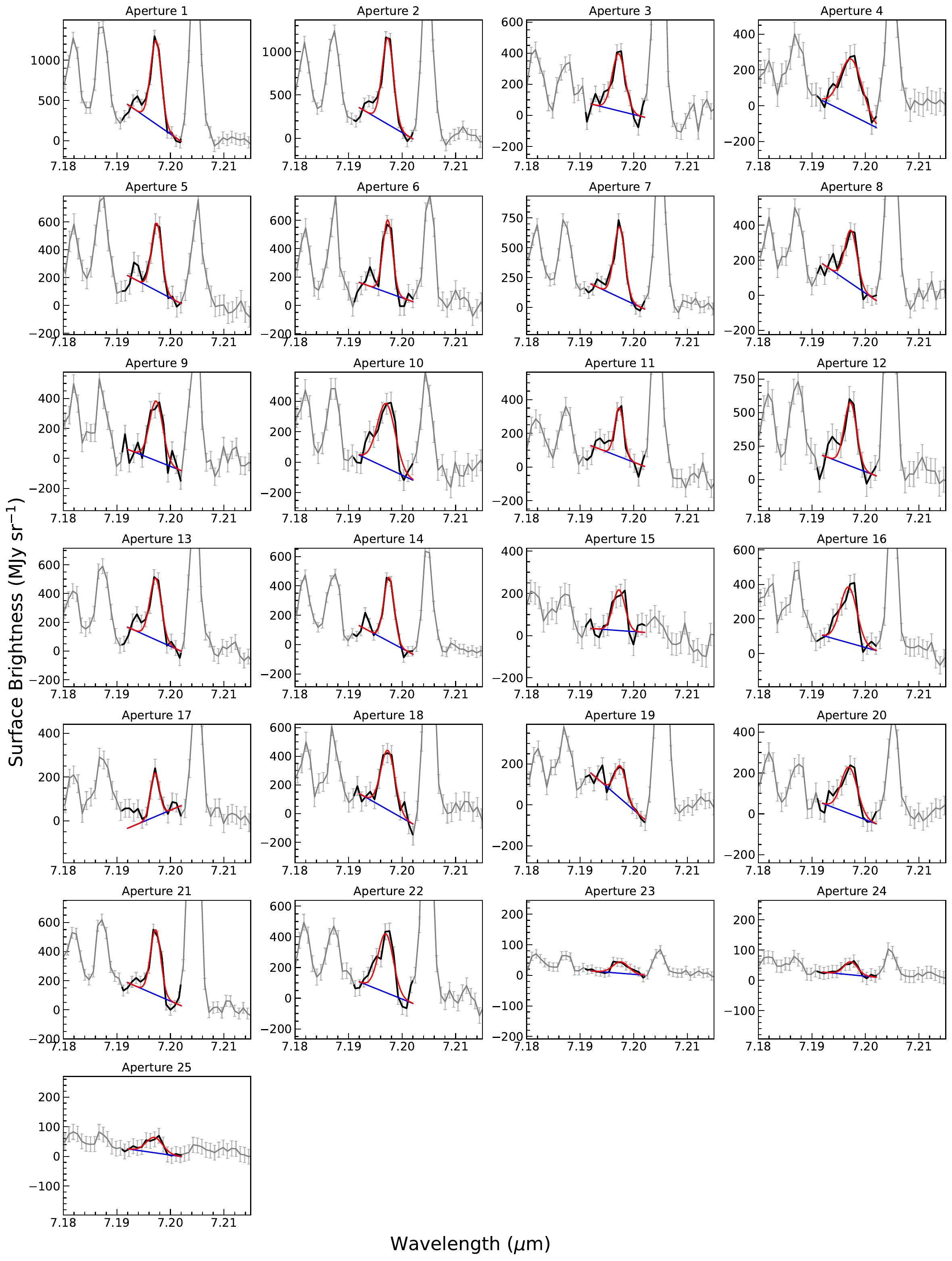}
    \caption{Gaussian fits to the 7.2 band used for the band ratio method to determine the temperature. The integrated surface brightness value at each aperture is listed in Table~\ref{tab:best_fit_results}. }
    \label{fig:gaussfits_7p2}
\end{figure*}

The best-fit Gaussians to the 7.15~\micron and 7.2~\micron bands for all 25 apertures are shown in Figs.~\ref{fig:gaussfits_7p15} and \ref{fig:gaussfits_7p2}, respectively. These are used to determine the excitation temperature of \chthreeplus using the band ratio method as described in Sec.~\ref{sec:detection-of-CH3plus}. 

\label{app_sec:band-ratio}

\section{Fit results}
\label{sec:fit_results}

The best-fit models for determining \chthreeplus temperature and column density in all 25 apertures are shown in Fig.~\ref{fig:bestfit_plots}. See Sec.~\ref{sec:detection-of-CH3plus} for details on the fitting procedure.

\begin{figure*}
  \gridline{\fig{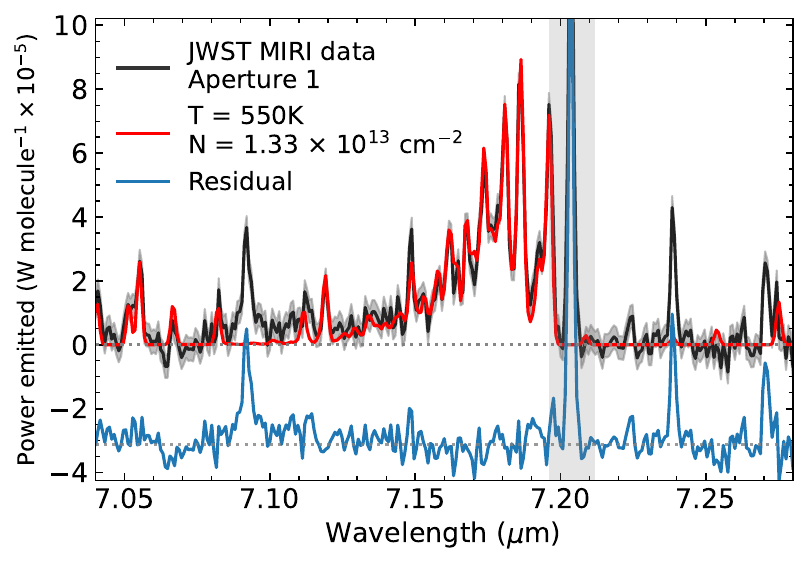}{0.45\textwidth}{(a) Aperture 1.}
            \fig{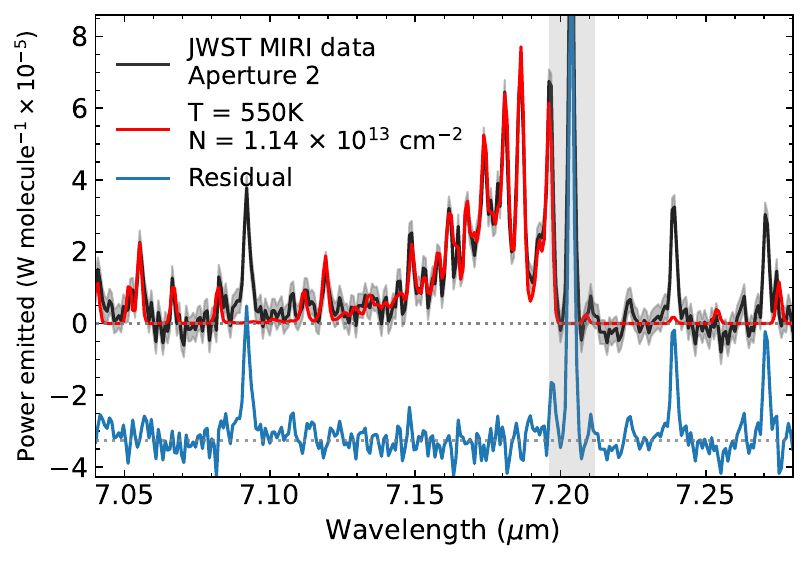}{0.45\textwidth}{(b) Aperture 2.}}
  \gridline{\fig{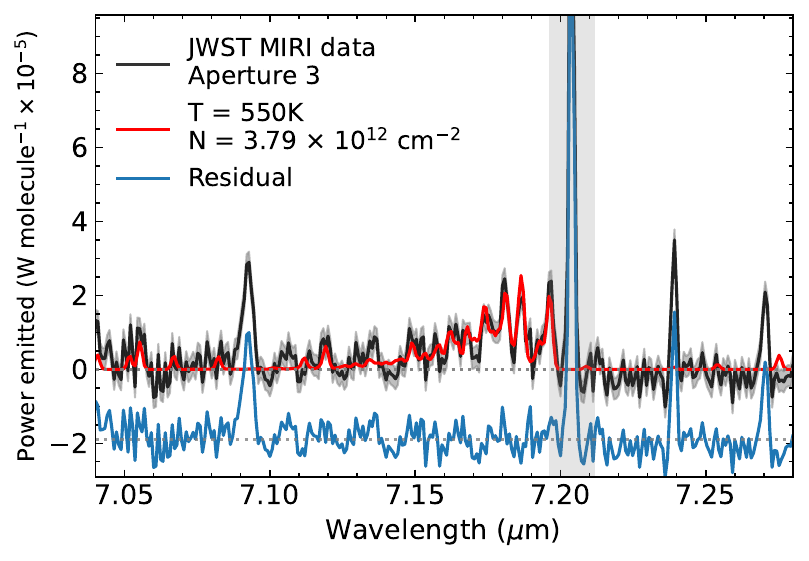}{0.45\textwidth}{(c) Aperture 3.}
            \fig{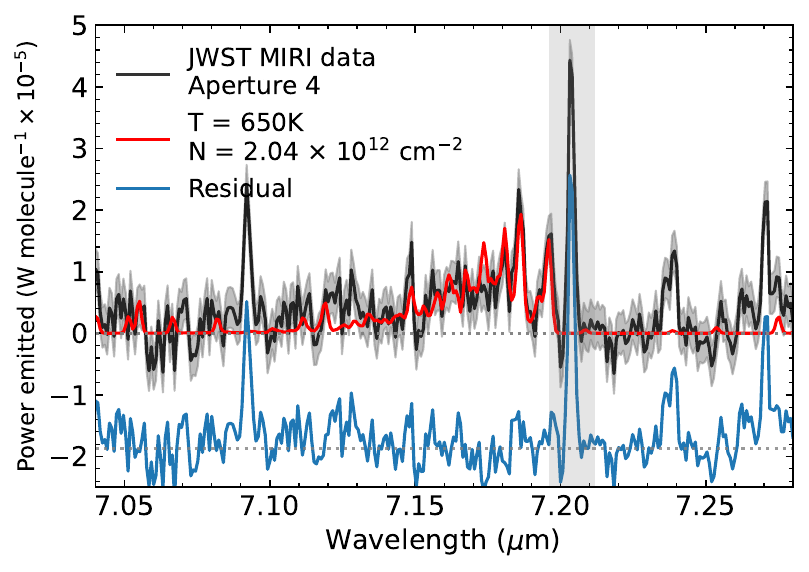}{0.45\textwidth}{(d) Aperture 4.}}
  \gridline{\fig{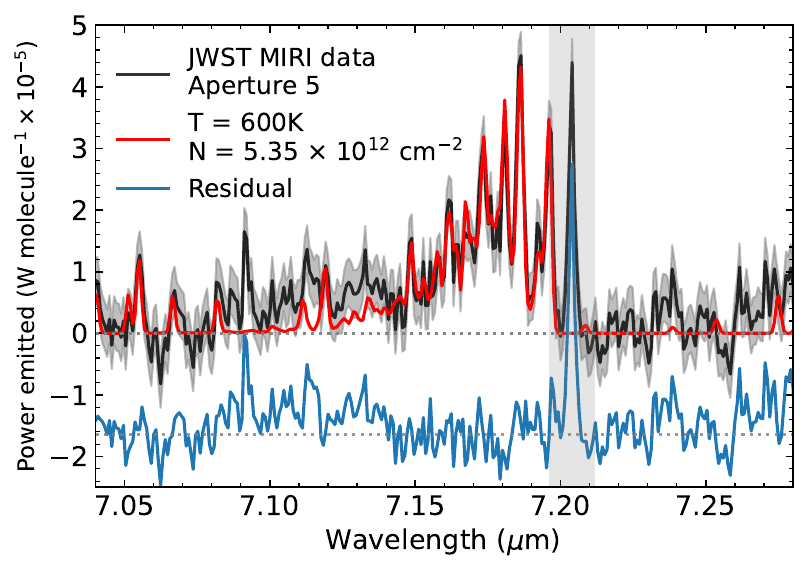}{0.45\textwidth}{(e) Aperture 5.}
            \fig{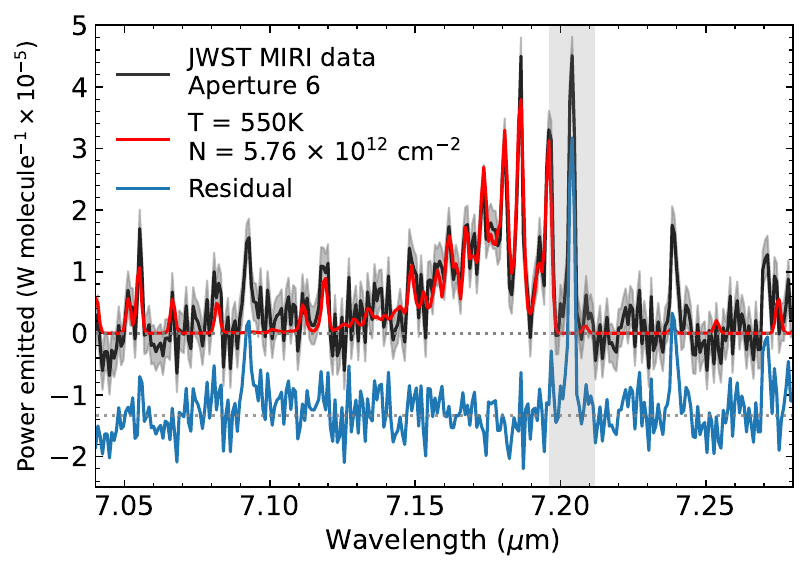}{0.45\textwidth}{(f) Aperture 6.}}

  \caption{Observations (black), best-fit model (red), and residuals (blue) for each aperture.}
  \label{fig:bestfit_plots}
\end{figure*}

\begin{figure*}
  \ContinuedFloat      
  \gridline{\fig{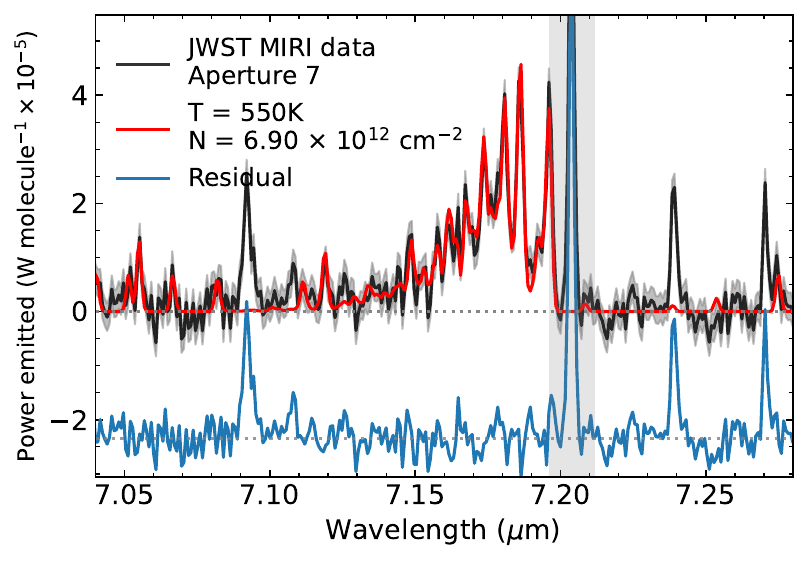}{0.45\textwidth}{(g) Aperture 7.}
            \fig{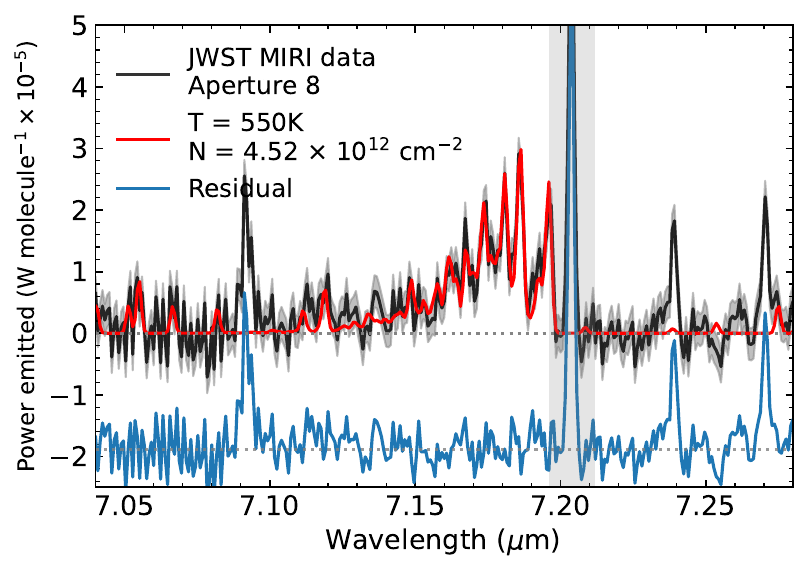}{0.45\textwidth}{(h) Aperture 8.}}
  \gridline{\fig{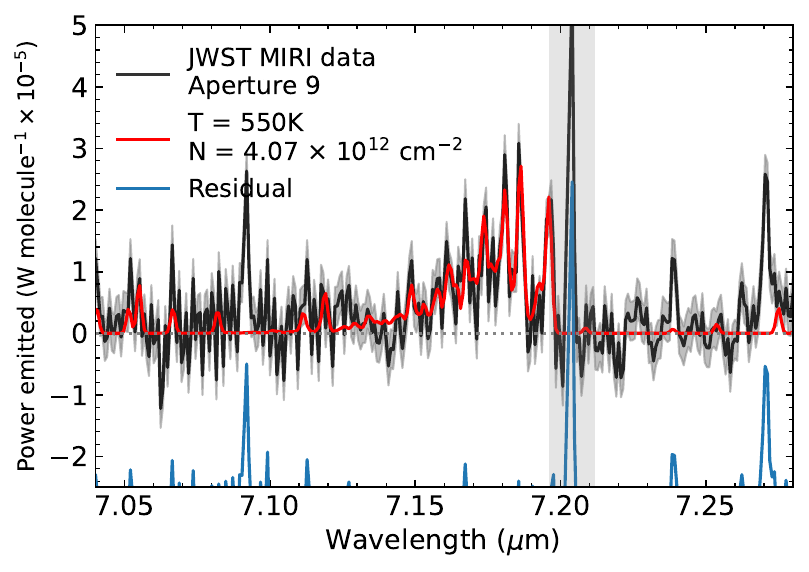}{0.45\textwidth}{(i) Aperture 9.}
            \fig{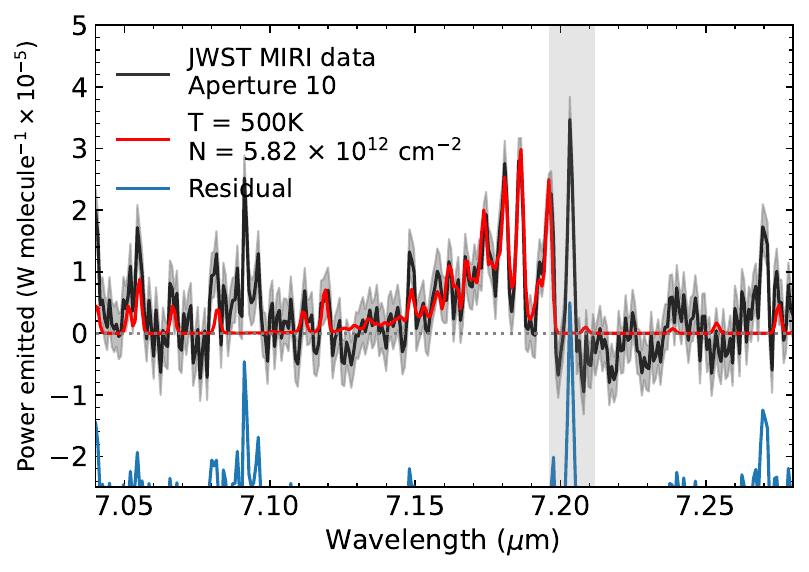}{0.45\textwidth}{(j) Aperture 10.}}
  \gridline{\fig{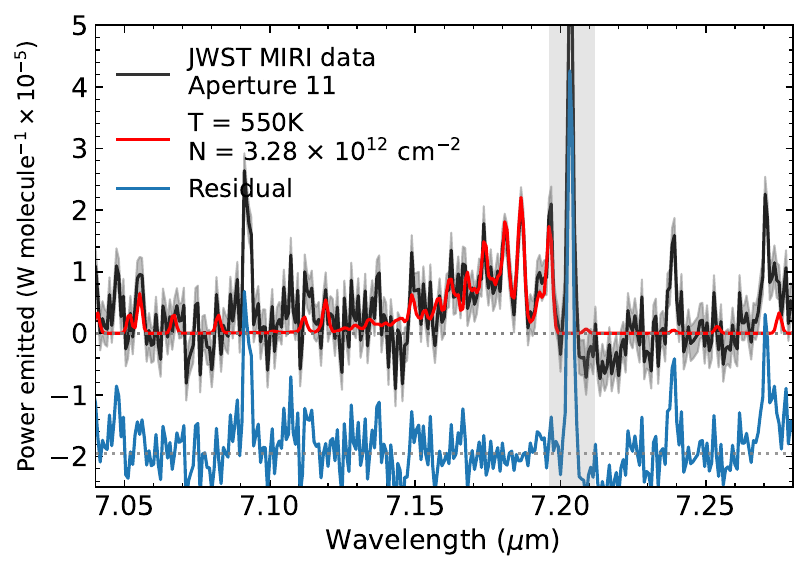}{0.45\textwidth}{(k) Aperture 11.}
            \fig{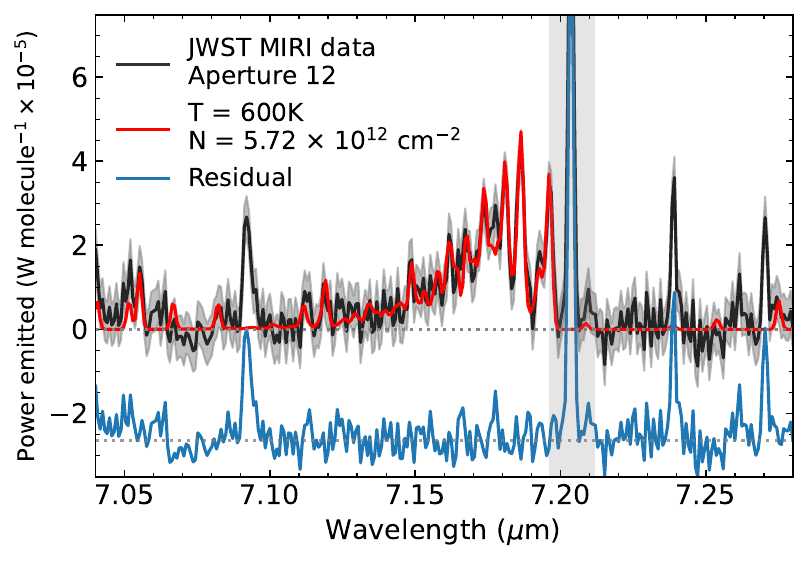}{0.45\textwidth}{(l) Aperture 12.}}

  \caption{Figure \ref{fig:bestfit_plots} (continued).}
\end{figure*}

\begin{figure*}
  \ContinuedFloat
  \gridline{\fig{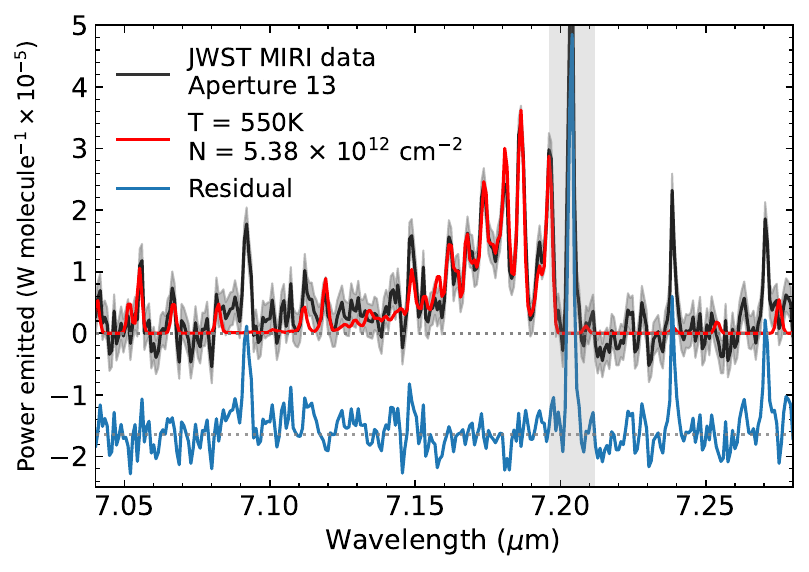}{0.45\textwidth}{(m) Aperture 13.}
            \fig{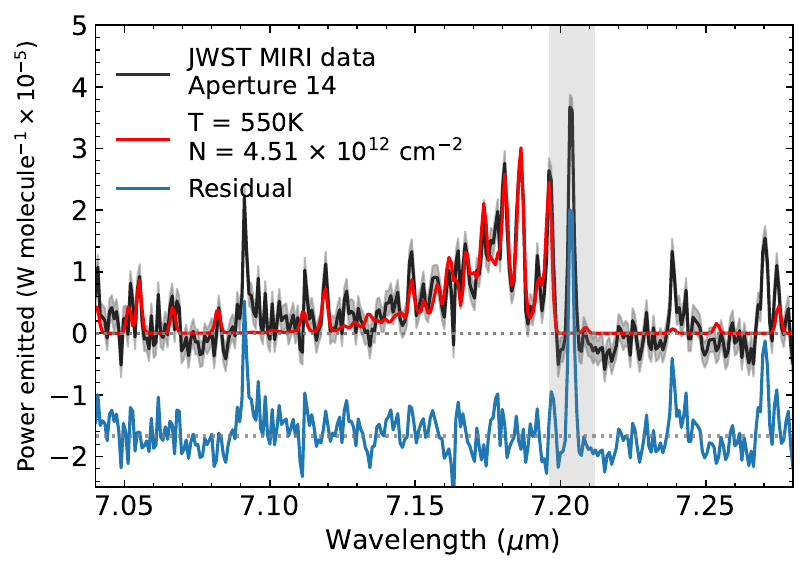}{0.45\textwidth}{(n) Aperture 14.}}
  \gridline{\fig{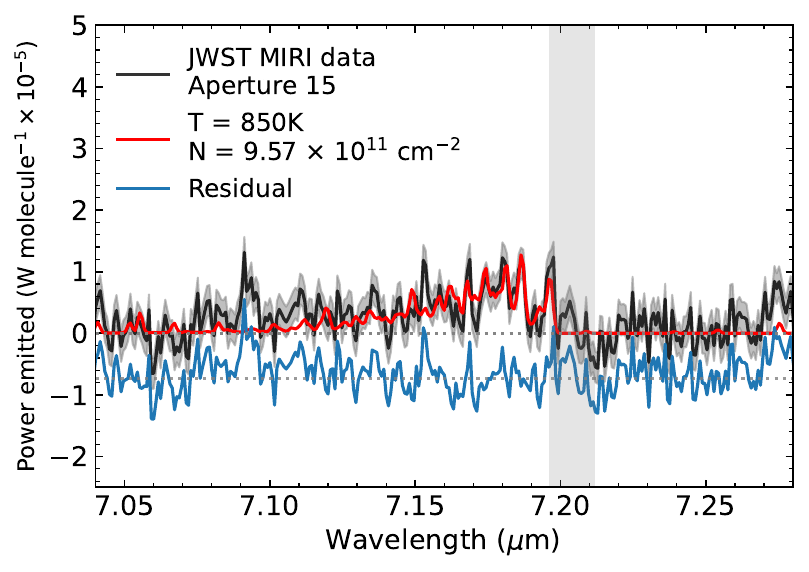}{0.45\textwidth}{(o) Aperture 15.}
            \fig{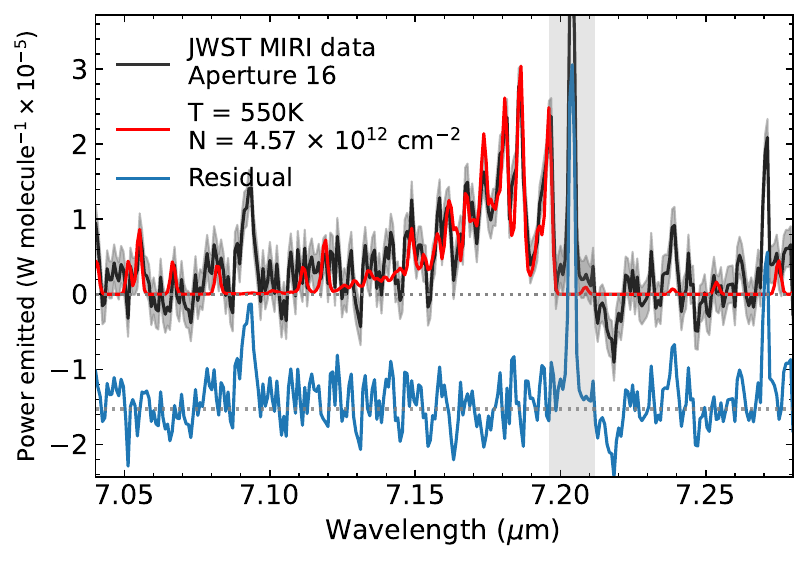}{0.45\textwidth}{(p) Aperture 16.}}
  \gridline{\fig{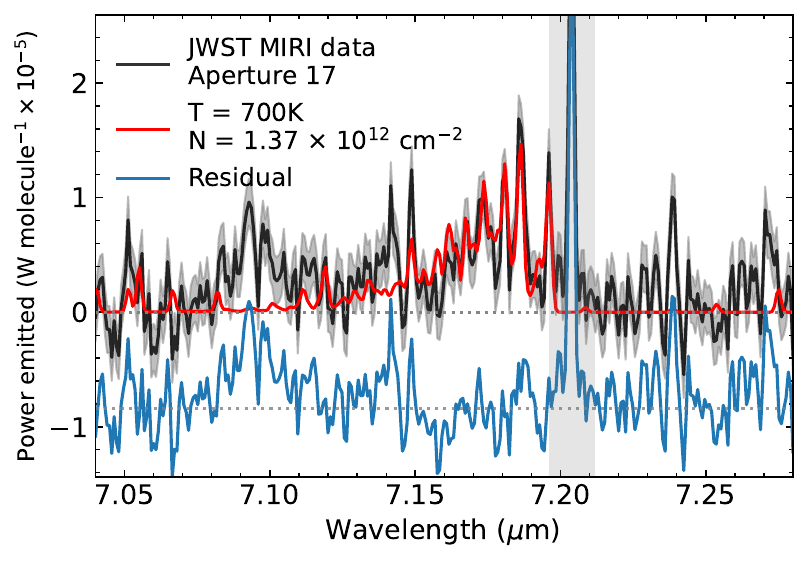}{0.45\textwidth}{(q) Aperture 17.}
            \fig{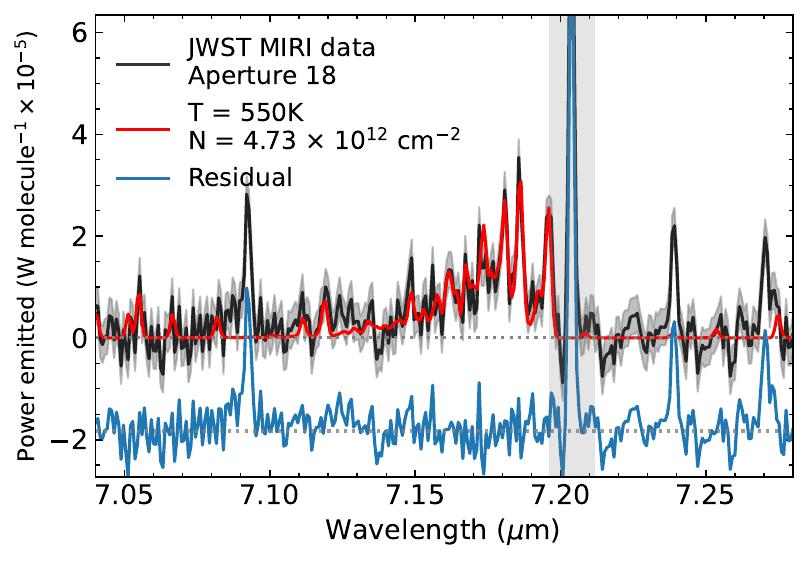}{0.45\textwidth}{(r) Aperture 18.}}

  \caption{Figure \ref{fig:bestfit_plots} (continued).}
\end{figure*}

\begin{figure*}
  \ContinuedFloat
  \gridline{\fig{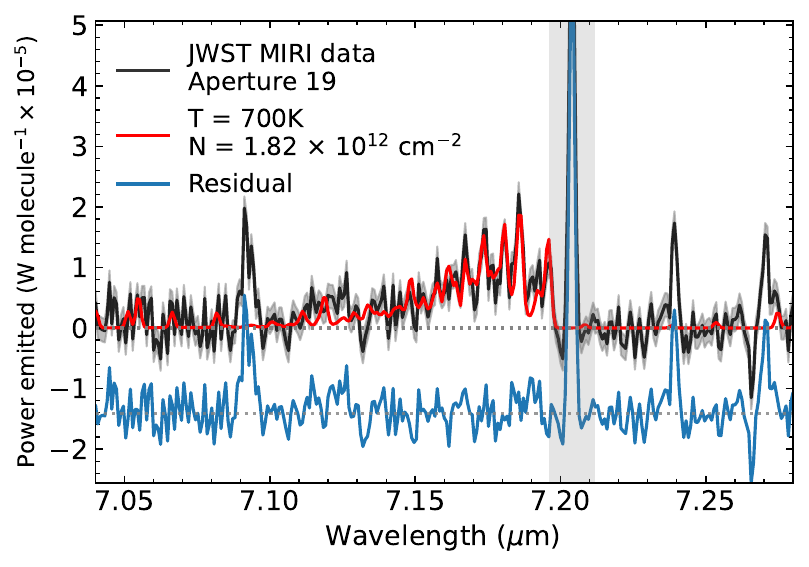}{0.45\textwidth}{(s) Aperture 19.}
            \fig{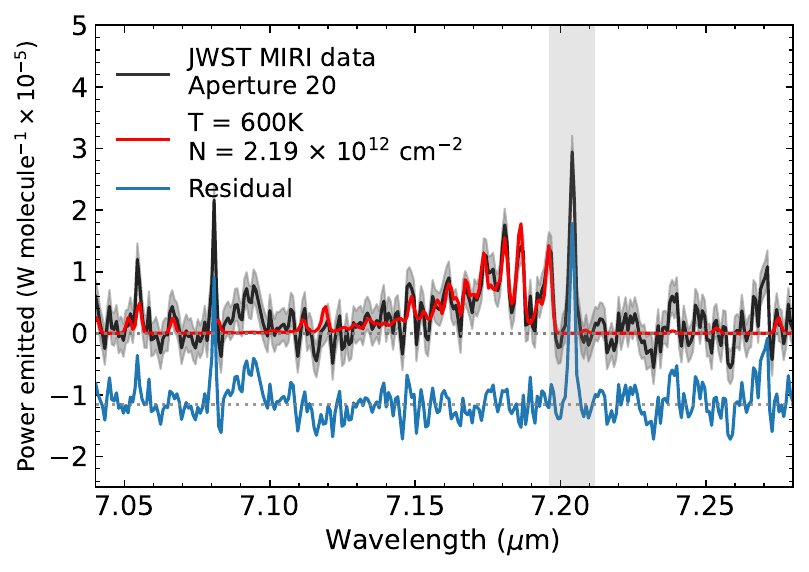}{0.45\textwidth}{(t) Aperture 20.}}
  \gridline{\fig{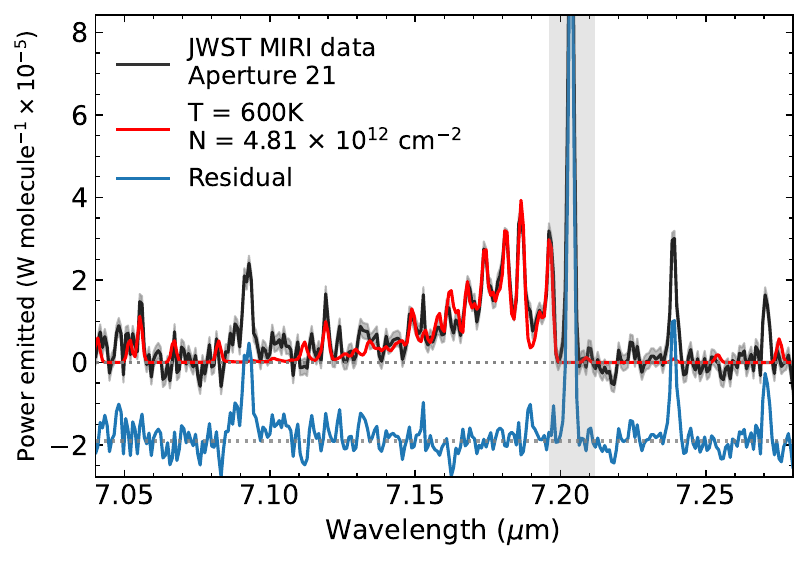}{0.45\textwidth}{(u) Aperture 21.}
            \fig{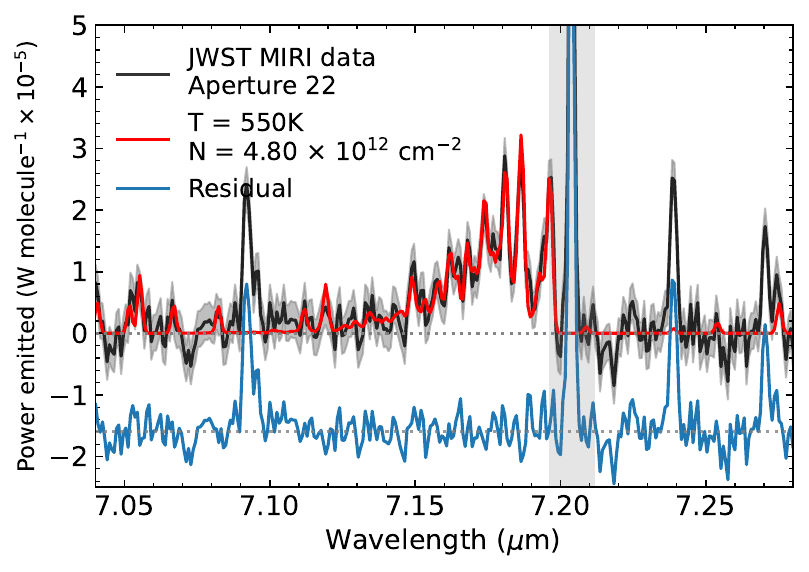}{0.45\textwidth}{(v) Aperture 22.}}
  \gridline{\fig{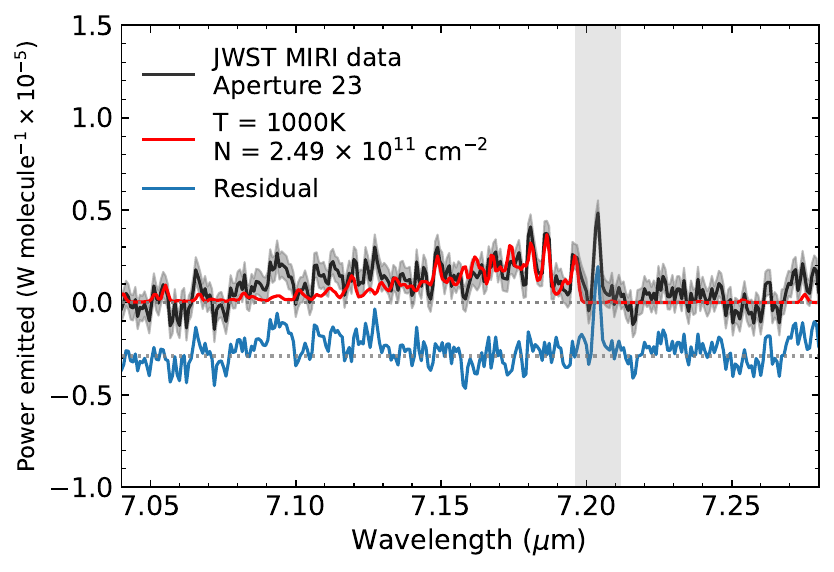}{0.45\textwidth}{(u) Aperture 23.}
            \fig{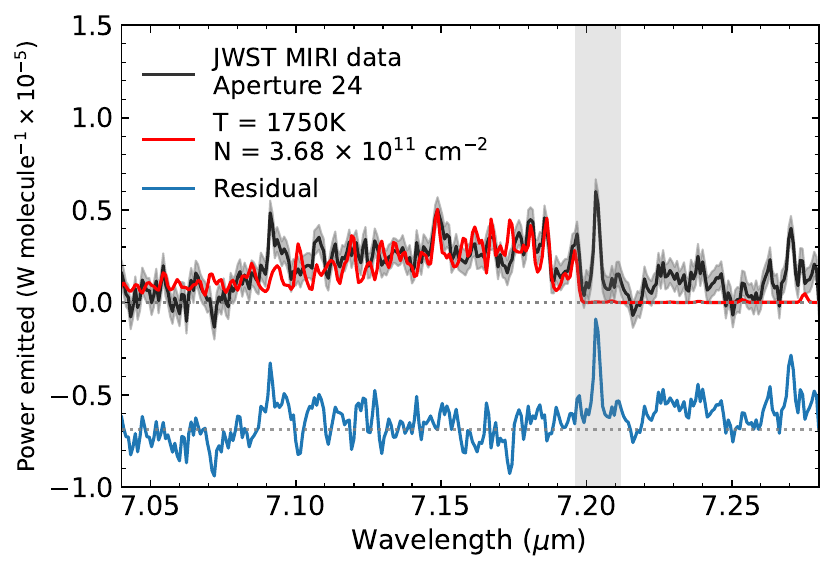}{0.45\textwidth}{(v) Aperture 24.}}

  \caption{Figure \ref{fig:bestfit_plots} (continued).}
\end{figure*}

\begin{figure*}
  \ContinuedFloat
  \gridline{\fig{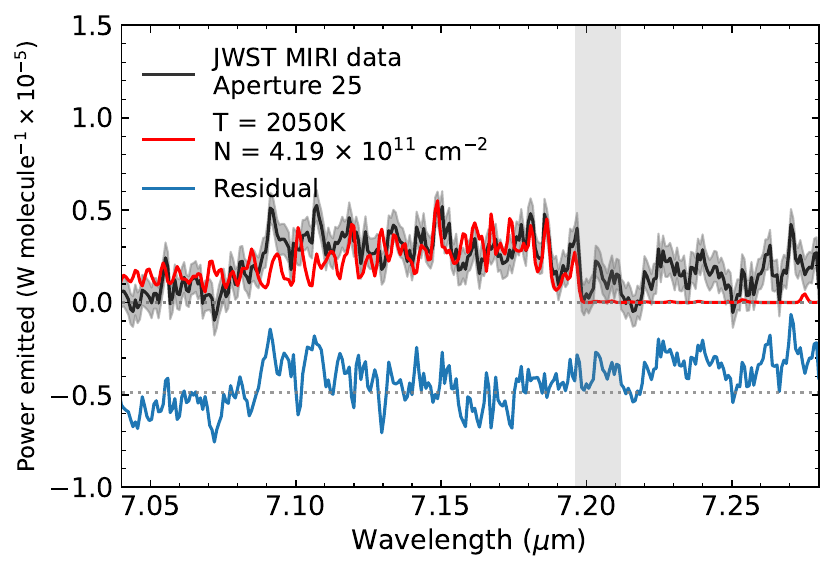}{0.45\textwidth}{(s) Aperture 25.}
            }

  \caption{Figure \ref{fig:bestfit_plots} (continued).}
\end{figure*}

\clearpage
\bibliography{ALL}{}
\bibliographystyle{aasjournalv7}

\end{document}